\documentclass[letterpaper,twocolumn,10pt]{article}
\usepackage{usenix2019_v3}
\usepackage{hyperref}

\newcommand{\attackName}{Spoiler}
\newcommand{\attack}{\textsc{\attackName}}

\title{\Large \bf \attack:
Speculative Load Hazards Boost Rowhammer and Cache Attacks
}

\usepackage{authblk}

\makeatletter

 \author[1]{Saad Islam}
 \author[1]{Ahmad Moghimi}
 \author[2]{Ida Bruhns}
 \author[2]{Moritz Krebbel}
 \author[1]{Berk Gulmezoglu}
 \author[1, 2]{Thomas Eisenbarth}
 \author[1]{Berk Sunar}

 \affil[1]{Worcester Polytechnic Institute, Worcester, MA, USA}
 \affil[2]{ University of L\"ubeck, L\"ubeck, Germany}

\usepackage{comment}
\usepackage{makecell}
\usepackage{listings}
\usepackage{color}
\usepackage{algpseudocode}
\usepackage{algorithmicx}
\usepackage[ruled]{algorithm}
\usepackage{algpseudocode} 
\usepackage{amsmath}
\usepackage{graphicx}
\usepackage{subfigure}
\usepackage{amssymb}
\usepackage{epstopdf}
\usepackage{tikz} 
\usetikzlibrary{arrows, automata, positioning}
\usepackage{environ}
\usepackage[colorinlistoftodos,prependcaption,textsize=tiny]{todonotes}
\usepackage{enumitem}

    \definecolor{linkcolor}{rgb}{0.65,0,0}
    \definecolor{citecolor}{rgb}{0,0.4,0}
    \definecolor{urlcolor}{rgb}{0,0,0.65}
    \hypersetup{pdfpagemode=UseNone,pdfstartview=FitH,colorlinks=true, linkcolor=linkcolor, urlcolor=urlcolor, citecolor=citecolor}

\definecolor{dkgreen}{rgb}{0,0.6,0}
\definecolor{gray}{rgb}{0.5,0.5,0.5}
\definecolor{mauve}{rgb}{0.58,0,0.82}
\begin{document}
\maketitle

\pagestyle{empty}

\begin{abstract}
    Modern microarchitectures incorporate optimization techniques such as \textit{speculative loads} and \textit{store forwarding} to improve the memory bottleneck. The processor executes the \texttt{load} speculatively before the \texttt{stores}, and forwards the data of a preceding \texttt{store} to the \texttt{load} if there is a potential dependency. This enhances performance since the \texttt{load} does not have to wait for preceding \texttt{stores} to complete. However, the dependency prediction relies on partial address information, which may lead to false dependencies and stall hazards. 
    
    In this work, we are the first to show that the dependency resolution logic that serves the speculative load can be exploited to gain information about the physical page mappings. Microarchitectural side-channel attacks such as Rowhammer and cache attacks like Prime+Probe rely on the reverse engineering of the virtual-to-physical address mapping. We propose the~\attack~attack which exploits this leakage to speed up this reverse engineering by a factor of 256. Then, we show how this can improve the Prime+Probe attack by a 4096 factor speed up of the eviction set search, even from sandboxed environments like JavaScript. Finally, we improve the Rowhammer attack by showing how \attack\ helps to conduct DRAM row conflicts deterministically with up to 100\% chance, and by demonstrating a double-sided Rowhammer attack with normal user's privilege. The later is due to the possibility of detecting contiguous memory pages using the \attack\ leakage. 
\end{abstract}



\section{Introduction}
Microarchitectural attacks have evolved over the past decade from attacks on weak cryptographic implementations~\cite{bernstein2005cache} to devastating attacks breaking through layers of defenses provided by the hardware and the Operating System (OS)~\cite{van2018foreshadow}. These attacks can steal secrets such as cryptographic keys~\cite{benger2014ooh,pereida2016make} or keystrokes~\cite{lipp2017practical}. More advanced attacks can entirely subvert the OS memory isolation to read the memory content from more privileged security domains~\cite{lipp2018meltdown}, and to bypass defense mechanisms such as Kernel Address Space Layout Randomization (KASLR)~\cite{gruss2016prefetch,evtyushkin2016jump}. Rowhammer attacks can further break the data and code integrity by tampering with memory contents~\cite{kim2014flipping,seaborn2015exploiting}. While most of these attacks require local access and native code execution, various efforts have been successful in conducting them remotely~\cite{tatar2018throwhammer} or from within a remotely accessible sandbox such as JavaScript~\cite{orenspy}.

Memory components such as DRAM~\cite{kim2014flipping} and cache~\cite{percival2005cache} are not the only microarchitectural attack surfaces. \textit{Spectre} attacks on the branch prediction unit~\cite{kocher2018spectre,maisuradze2018ret2spec} imply that side channels such as caches can be used as a primitive for more advanced attacks on speculative engines. Speculative engines predict the outcome of an operation before its completion, and they enable execution of the following dependent instructions ahead of time based on the prediction. As a result, the pipeline can maximize the instruction level parallelism and resource usage. In rare cases where the prediction is wrong, the pipeline needs to be flushed resulting in performance penalties. However, this approach suffers from a security weakness, in which an adversary can fool the predictor and introduce arbitrary mispredictions that leave microarchitectural footprints in the cache. These footprints can be collected through the cache side channel to steal secrets. 


Modern processors feature further speculative behavior such as \textit{memory disambiguation} and speculative loads~\cite{doweck2006inside}. A \texttt{load} operation can be executed speculatively before preceding \texttt{store} operations. During the speculative execution of the \texttt{load}, false dependencies may occur due to the unavailability of physical address information. These false dependencies need to be resolved to avoid computation on invalid data. The occurrence of false dependencies and their resolution depend on the actual implementation of the memory subsystem. Intel uses a proprietary memory disambiguation and \textit{dependency resolution logic} in the processors to predict and resolve false dependencies that are related to the speculative load. In this work, we discover that the dependency resolution logic suffers from an unknown false dependency independent of the 4K aliasing~\cite{moghimi2018memjam, sullivan2018microarchitectural}. The discovered false dependency happens during the 1\,MB aliasing of speculative memory accesses which is exploited to leak information about physical page mappings.

The state-of-the-art microarchitectural attacks~\cite{irazoqui2015s,pessl2016drama} either rely on knowledge of physical addresses or are significantly eased by that knowledge. Yet,
knowledge of the physical address space is only granted with root privileges. Cache attacks such as \textit{Prime+Probe} on the Last-Level Cache (LLC) are challenging due to the unknown mapping of virtual addresses to cache sets and slices. Knowledge about the physical page mappings enables more attack opportunities using the Prime+Probe technique. Rowhammer~\cite{kim2014flipping} attacks require efficient access to rows within the same bank to induce fast row conflicts. To achieve this, an adversary needs to reverse engineer layers of abstraction from the virtual address space to DRAM cells. Availability of physical address information facilitates this reverse engineering process. In sandboxed environments, attacks are more limited, since in addition to the limited access to the address space, low-level instructions are also inaccessible~\cite{gruss2016rowhammer}. Previous attacks assume special access privileges only granted through weak software configurations~\cite{irazoqui2015s,lipp2016armageddon,van2016drammer} to overcome some of these challenges. In contrast, \attack\ only relies on simple operations, \texttt{load} and \texttt{store}, to recover crucial physical address information, which in turn enables Rowhammer and cache attacks, by leaking information about physical pages without assuming any weak configuration or special privileges. 

\subsection{Our Contribution}
We have discovered a novel microarchitectural leakage which reveals critical information about physical page mappings to user space processes. The leakage can be exploited by a limited set of instructions, which is visible in all Intel generations starting from the $1^{st}$ generation of Intel Core processors, independent of the OS and also works from within virtual machines and sandboxed environments. In summary, this work: 
\begin{enumerate}
    \item exposes a previously unknown microarchitectural leakage stemming from the false dependency hazards during speculative load operations.

    \item proposes an attack,~\attack, to efficiently exploit this leakage to speed up the reverse engineering of virtual-to-physical mappings by a factor of 256 from both native and JavaScript environments.

    \item demonstrates a novel eviction set search technique from JavaScript and compares its reliability and efficiency to existing approaches.
    \item 
    achieves efficient DRAM row conflicts and the first \textit{double-sided Rowhammer} attack with normal user-level privilege using the contiguous memory detection capability of \attack.
    
    \item explores how \attack\ can track nearby load operations from a more privileged security domain right after a context switch. 
    
\end{enumerate}

\subsection{Related Work}
Kosher et al. \cite{kocher2018spectre} and Maisuradze et al. \cite{maisuradze2018ret2spec} have exploited vulnerabilities in the speculative branch prediction unit. Transient execution of instructions after a fault, as exploited by Lipp et al. \cite{lipp2018meltdown} and Bulck et al. \cite{van2018foreshadow}, can leak memory content of protected environments. Similarly, transient behavior due to the lazy store/restore of the FPU and SIMD registers can leak register contents from other contexts~\cite{stecklina2018lazyfp}. New variants of both Meltdown and Spectre have been systematically analyzed~\cite{canella2018systematic}. The Speculative Store Bypass (SSB) vulnerability~\cite{SSB} is a variant of the Spectre attack and relies on the stale sensitive data in registers to be used as an address for speculative loads which may then allow the attacker to read this sensitive data. In contrast to previous attacks on speculative and transient behaviors, we discover a new leakage on the undocumented memory disambiguation and dependency resolution logic. \attack\ is {\bf not} a Spectre attack. The root cause for \attack\ is a weakness in the address speculation of Intel's proprietary implementation of the memory subsystem which directly leaks timing behavior due to physical address conflicts. Existing spectre mitigations would therefore not interfere with \attack.

The timing behavior of the 4K aliasing false dependency on Intel processors have been studied~\cite{agneroptimize,yarom2017cachebleed}. \textit{MemJam}~\cite{moghimi2018memjam} uses this behavior to perform a side-channel attack, and Sullivan et al.~\cite{sullivan2018microarchitectural} demonstrate a covert channel. These works only mention the 4K aliasing as documented by Intel~\cite{inteloptimze}, and the authors conclude that the address aliasing check is a two stage approach: Firstly, it uses page offset for the initial guess. Secondly, it performs the final resolution based on the exact physical address. On the contrary, we discover that the undocumented \textit{address resolution} logic performs additional partial address checks that lead to an unknown, but observable aliasing behavior based on the physical address.

Several microarchitectural attacks have been discovered to recover virtual address information and break KASLR by exploiting the Translation Lookaside Buffer (TLB)~\cite{hund2013practical}, Branch Target Buffer (BTB)~\cite{evtyushkin2016jump} and Transactional Synchronization Extensions (TSX)~\cite{jang2016breaking}. Additionally, Gruss et al.~\cite{gruss2016prefetch} exploit the timing information obtained from the \texttt{prefetch} instruction to leak the physical address information. The main obstacle to this approach is that the \texttt{prefetch} instruction is not accessible in JavaScript, and it can be disabled in native sandboxed environments~\cite{yee2009native}, whereas~\attack~is applicable to sandboxed environments including JavaScript. 

Knowledge of the physical address enables adversaries to bypass OS protections~\cite{kemerlis2014ret2dir} and ease other microarchitectural attacks~\cite{lipp2016armageddon}. For instance, the \texttt{procfs} filesystem exposes physical addresses~\cite{lipp2016armageddon}, and \textit{Huge pages} allocate contiguous physical memory~\cite{liu2015last,irazoqui2015s}. \textit{Drammer}~\cite{van2016drammer} exploits the Android \textit{ION memory allocator} to access contiguous memory. However, access to the aforementioned primitives is restricted on most environments by default. We do not have any assumption about the OS and software configuration, and we exploit a hardware leakage with minimum access rights to find virtual pages that have the same least significant 20 physical address bits. \textit{GLitch}~\cite{frigo2018grand} detects contiguous physical pages by exploiting row conflicts through the GPU interface. In contrast, our attack does not rely on a specific integrated GPU configuration, and it is widely applicable to any system running on an Intel CPU. We use \attack\ to find contiguous physical pages with a high probability and verify it by producing row conflicts. \attack\ is particularly helpful for attacks in sandboxed low-privilege environments such as JavaScript, where previous methods require a time-consuming brute forcing of the memory addresses~\cite{orenspy,seaborn2015exploiting,gruss2016rowhammer}.

\section{Background}
\subsection{Memory Management} \label{sec:memman}
The virtual memory manager shares the DRAM across all running tasks by assigning isolated virtual address spaces to each task. The assigned memory is allocated in pages, which are typically 4\,kB each, and each virtual page will be stored as a physical page in DRAM through a virtual-to-physical page mapping. Memory instructions operate on virtual addresses, which are translated within the processor to the corresponding physical addresses. The page offset comprising the least significant 12 bits of the virtual address is not translated. The processor only translates the bits in the rest of the virtual address, the virtual page number. The OS is the reference for this translation, and the processor stores the translation results inside the TLB. As a result, repeated translations of the same address are performed more efficiently. 

\subsection{Cache Hierarchy}
Modern processors incorporate multiple levels of caches to avoid the DRAM access latency. The cache memory on Intel processors is organized into sets and slices. Each set can store a certain number of lines, where the line size is 64 bytes. The 6 Least Significant Bits (LSBs) of the physical address are used to determine the offset within a line and the remaining bits are used to determine which set to store the cache line in. The number of physical address bits that are used for mapping is higher for the LLC, since it has a large number of sets, e.g., 8192 sets. Hence, the untranslated part of the virtual address bits which is the page offset, cannot be used to index the LLC sets. Instead, higher physical address bits are used. Further, each set of LLC is divided into multiple slices, one slice for each logical processor. The mapping of the physical addresses to the slices uses an undocumented function~\cite{irazoqui2015systematic}. When the processor accesses a memory address, a cache hit or miss occurs. If a miss occurs in all cache levels, the memory line has to be fetched from DRAM. Accesses to the same memory address would be served from the cache unless other memory accesses evict that cache line. In addition, we can use the \texttt{clflush} instruction, which follows the same memory access check as other memory operations, to evict our own cache lines from the entire cache hierarchy.

\subsection{Prime+Probe Attack}

In the Prime+Probe attack, the attacker first fills an entire cache set by accessing memory addresses that are mapped to the same set, an \textit{eviction set}. Later, the attacker checks whether the victim program has displaced any entry in the cache set by accessing the eviction set again and measuring the execution time. If this is the case, the attacker can detect congruent addresses, since the displaced entries cause an increased access time. However, finding the eviction sets is difficult due to the unknown translation of virtual addresses to physical addresses. Since an unprivileged attacker has no access to hugepages~\cite{inci2016cache} or the virtual-to-physical page mapping such as the \texttt{pagemap} file~\cite{lipp2016armageddon}, knowledge about the physical address bits greatly speeds up the eviction set search.

\subsection{Rowhammer Attack} \label{sec:backrow}

DRAM consists of multiple memory banks, and each bank is subdivided into rows. When the processor accesses a memory location, the corresponding row needs to be activated and loaded into the row buffer. If the processor accesses the same row again, it is called a row hit, and the request will be served from the row buffer. Otherwise, it is called a row conflict, and the previous row will be deactivated and copied back to the original row location, after which the new row is activated. DRAM cells leak charge over time and need to be refreshed periodically to maintain the data. A Rowhammer~\cite{kim2014flipping} attack causes cells of a victim row to leak faster by activating the neighboring rows repeatedly. If the refresh cycle fails to refresh the victim row fast enough, that leads to bit flips. Once bit flips are found, they can be exploited by placing any security-critical data structure or code page at that particular location and triggering the bit flip again~\cite{xiao2016one,seaborn2015exploiting,gruss2018another}. The Rowhammer attack requires fast access to the same DRAM cells by bypassing the CPU cache, e.g., using \texttt{clflush}~\cite{kim2014flipping}. Additionally, cache eviction based on an eviction set can also result in access to DRAM cells when \texttt{clflush} is not available~\cite{gruss2016rowhammer,aweke2016anvil}. Efficiently building eviction sets may thus also enhance Rowhammer attacks. For a successful Rowhammer attack, it is essential to collocate multiple memory pages within the same bank and adjacent to each other. A number of physical address bits, depending on the hardware configuration, are used to map memory pages to banks~\cite{pessl2016drama}. Since the rows are generally placed sequentially within the banks, access to adjacent rows within the same bank can be achieved if we have access to contiguous physical pages.

\subsection{Memory Order Buffer} \label{sec:mob}
\begin{figure}[t!]
  \includegraphics[width=\linewidth]{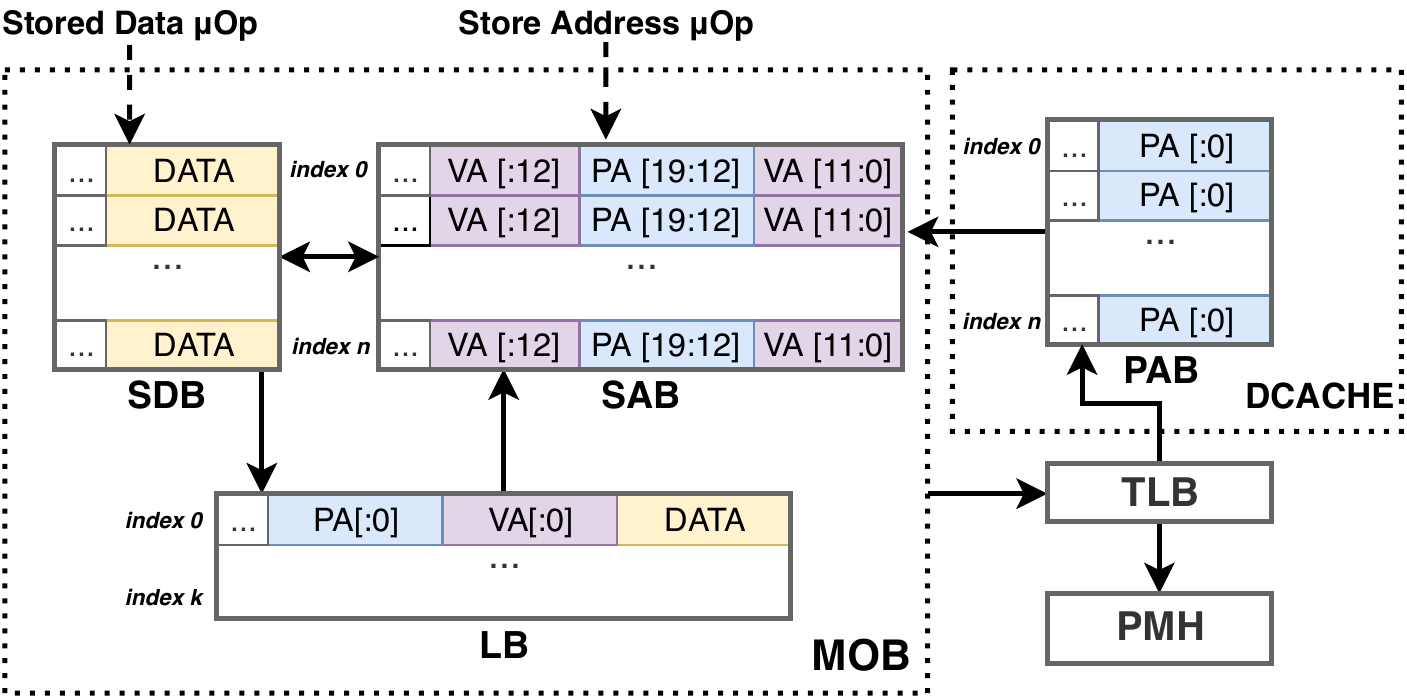}
  \caption{The Memory Order Buffer includes circular buffers SDB, SAB and LB. SDB, SAB and PAB of the DCACHE have the same number of entries. SAB may initially hold the virtual address and the partial physical address. MOB requests the TLB to translate the virtual address and update the PAB with the translated physical address.}
  \label{fig:mob}
  \vspace{-2.7ex}
\end{figure}

The processor manages memory operations using the Memory Order Buffer (MOB). MOB is tightly coupled with the data cache. The MOB assures that memory operations are executed efficiently by following the Intel memory ordering rule~\cite{IntelOrdering} in which memory \texttt{stores} are executed in-order and memory \texttt{loads} can be executed out-of-order. These rules have been enforced to improve the efficiency of memory accesses, while guaranteeing their correct commitment. \autoref{fig:mob} shows the MOB schematic according to Intel~\cite{abramson1998method,abramson2002method}. The MOB includes circular buffers, \textit{store buffer}\footnote{Store buffer consists of \textit{Store Address Buffer (SAB)} and \textit{Store Data Buffer (SDB)}. For simplicity, we use \textit{Store Buffer} to mention the logically combined SAB and SDB units. 
} and \textit{load buffer} (LB). A \texttt{store} will be decoded into two micro ops to store the address and data, respectively, to the store buffer. The store buffer enables the processor to continue executing other instructions before commitment of the \texttt{stores}. As a result, the pipeline does not have to stall for the \texttt{stores} to complete. This further enables the MOB to support out-of-order execution of the \texttt{load}. 

Store forwarding is an optimization mechanism that sends the \texttt{store} data to a \texttt{load} if the \texttt{load} address matches any of the store buffer entries. This is a speculative process, since the MOB cannot determine the true dependency of the \texttt{load} on \texttt{stores} based on the store buffer. Intel's implementation of the store buffer is undocumented, but a potential design suggests that it will only hold the virtual address, and it may include part of the physical address~\cite{abramson1998method,abramson2002method,kosinski2016store}. As a result, the processor may falsely forward the data, although the physical addresses do not match. The complete resolution will be delayed until the \texttt{load} commitment, since the MOB needs to ask the TLB for the complete physical address information, which is time consuming. Additionally, the data cache (DCACHE) may hold the translated store addresses in a Physical Address Buffer (PAB) with equal number of entries as the store buffer.

\section{Speculative Load Hazards} \label{sec:loadhazard}

\begin{figure*}[t!]
    \centering
    \includegraphics[width=.95\linewidth]{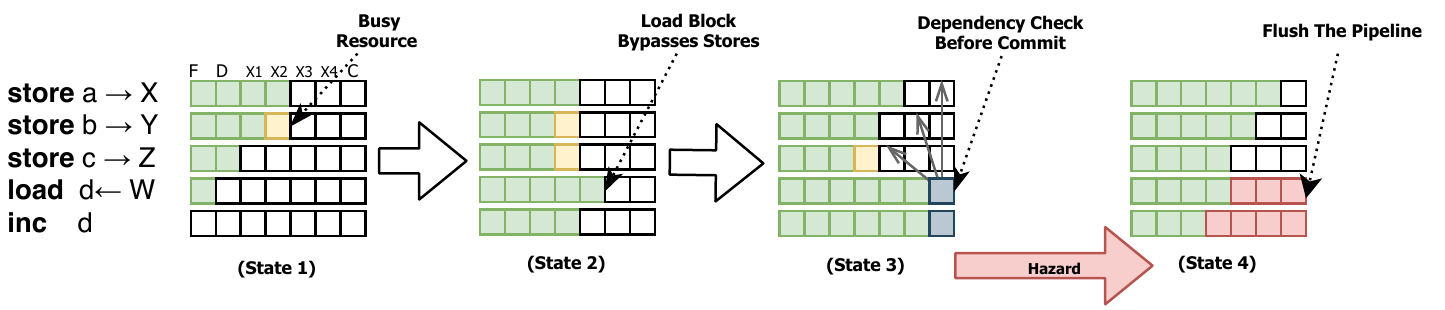}
    \caption{The speculative load is demonstrated on a hypothetical processor with 7 pipeline stages: $F$ = Fetch, $D$ = Decode, $X_{1-4}$ = Executions, and $C$ = Commit. When the memory \texttt{stores} are blocked competing for resources (State 1), the \texttt{load} will bypass the \texttt{stores} (State 2). The load block including the dependent instructions will not be committed until the dependency of the address $W$ versus $X$,$Y$,$Z$ are resolved (State 3). In case of a dependency hazard (State 4), the pipeline is flushed and the \texttt{load} is restarted.}
    \label{fig:loadhazard}
\end{figure*}

As we mentioned earlier, memory \texttt{loads} can be executed out-of-order and before the preceding memory \texttt{stores}. If one of the preceding \texttt{stores} modifies the content of a location in memory, the memory \texttt{load} address is referring to, out-of-order execution of the \texttt{load} will operate on stale data, which results in invalid execution of a program. This out-of-order execution of the memory \texttt{load} is a speculative behavior, since there is no guarantee during the execution time of the \texttt{load} that the virtual addresses corresponding to the memory \texttt{stores} do not conflict with the \texttt{load} address after translation to physical addresses. \autoref{fig:loadhazard} demonstrates this effect on a hypothetical processor with $7$ pipeline stages. As multiple \texttt{stores} may be blocked due to limited resources, the execution of the \texttt{load} and dependent instructions in the pipeline, the \emph{load block}, will bypass the \texttt{stores} since the MOB assumes the load block to be independent of the \texttt{stores}. This speculative behavior improves the memory bottleneck by letting other instructions continue their execution. However, if the dependency of the \texttt{load} and preceding \texttt{stores} is not verified, the load block may be computed on incorrect data which is either falsely forwarded by store forwarding (false dependency), or loaded from a stale cache line (unresolved true dependency). If the processor detects a false dependency before committing the \texttt{load}, it has to flush the pipeline and re-execute the load block. This will cause observable performance penalties and timing behavior. 

\subsection{Dependency Resolution}

Dependency checks and resolution occur in multiple stages depending on the availability of the address information in the store buffer. A \texttt{load} instruction needs to be checked against all preceding \texttt{stores} in the store buffer to avoid false dependencies and to ensure the correctness of the data. A potential design~\cite{hily2009resolving,kosinski2016store},\footnote{The implementation of the MOB used in Intel processors is unpublished and therefore we cannot be certain about the precise architecture. Our results agree with some of the possible designs that are described in the Intel patents.} suggests the following stages for the dependency check and resolution, as shown in~\autoref{fig:deplogic}: 

\begin{figure}[tp]
  \includegraphics[width=\linewidth]{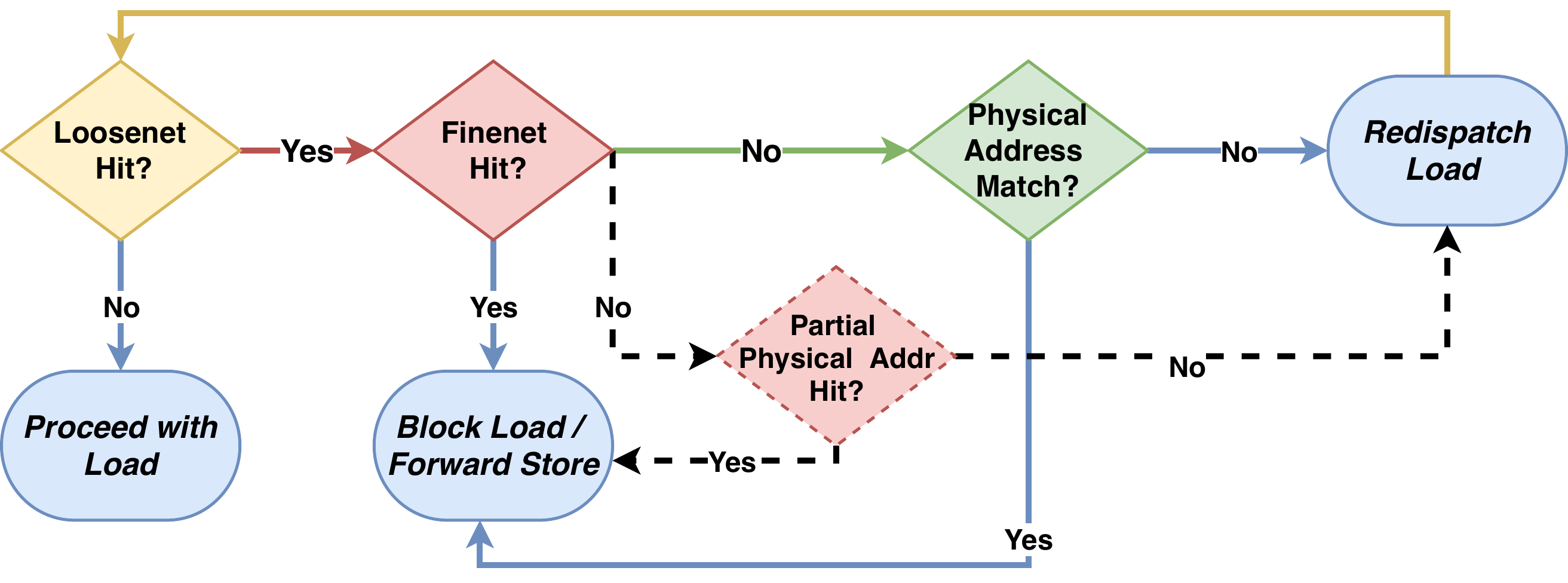}
  \caption{The dependency check logic: \textit{loosenet} initially checks the least 12 significant bits (page offset) and the \textit{finenet} checks the upper address bits, related to the page number. The final dependency using the physical address matching might still fail due to partial physical address checks.}
  \label{fig:deplogic}
\end{figure}

\begin{enumerate}
    \item \textbf{Loosenet}: The first stage is the \textit{loosenet} check where the page offsets of the \texttt{load} and \texttt{stores} are compared\footnote{According to \texttt{Ld\_Blocks\_Partial:Address\_Alias} Hardware Performance Counter (HPC) event\cite{inteloptimze}, loosenet is defined by Intel as the mechanism that only compare the page offsets.}. In case of a loosenet hit, the compared \texttt{load} and \texttt{store} may be dependent and the processor will proceed to the next check stage.

    \item \textbf{Finenet}: The next stage, called \textit{finenet}, uses upper address bits. The \textit{finenet} can be implemented to check the upper virtual address bits~\cite{hily2009resolving}, or the physical address tag~\cite{kosinski2016store}. Either way, it is an intermediate stage, and it is not the final dependency resolution. In case of a \textit{finenet} hit, the processor blocks the \texttt{load} and/or forwards the store data, otherwise, the dependency resolution will go into the final stage. 
    
    \item \textbf{Physical Address Matching}: At the final stage, the physical addresses will be checked. Since this stage is the final chance to resolve potential false dependencies, we expect the full physical address to be checked. However, one possible design suggests that if the physical addresses are not available, the physical address matching returns true and continues with the store forwarding~\cite{hily2009resolving}.

\end{enumerate}

\medskip\noindent
Since the page offset is identical between the virtual and physical address, loosenet can be performed as soon as the \texttt{store} is decoded. ~\cite{abramson1998method} suggests that the store buffer only holds bit 19 to 12 of the physical address. Although the PAB holds the full translated physical address, it is not clear in which stage this information can be available to the MOB. As a result, the \textit{finenet} check may be implemented based on checking the partial physical address bits. As we verify later, the dependency resolution logic may fail to resolve the dependency at multiple intermediate stages due to unavailability of the full physical address.

\section{The \attack~Attack}\label{sec:attack}

The attack model for \attack\ is the same as Rowhammer and cache attacks where the attacker's code is needed to be executed on the same underlying hardware as of the victim. 
As described in~\autoref{sec:loadhazard}, speculative loads may face other aliasing conditions in addition to the 4K aliasing, due to the partial checks on the higher address bits. To confirm this, we design an experiment to observe timing behavior of a speculative load based on higher address bits. For this purpose, we propose ~\autoref{alg:peaks} that executes a speculative load after multiple \texttt{stores} and further make sure to fill the store buffer with addresses that cause 4K aliasing during the execution of the \texttt{load}. Having $w$ as the window size, the algorithm iterates over a number of different memory pages, and for each page, it performs \texttt{stores} to that page and all previous $w$ pages within a \textit{window}. Since the size of the store buffer varies between different processor generations, we choose a big enough window ($w = 64$) to ensure that the \texttt{load} has 4K aliasing with the maximum number of entries in the store buffer and hence maximum potential conflicts. Following the \texttt{stores}, we measure the timing of a \texttt{load} operation from a different memory page, as defined by $x$. Since we want the \texttt{load} to be executed speculatively, we can not use a store fence such as \texttt{mfence} before the \texttt{load}. As a result, our measurements are an estimate of execution time for the speculatively \texttt{load} and nearby microarchitectural events. This may include a negligible portion of overhead for the execution of \texttt{stores}, and/or any delay due to the dependency resolution. If we iterate over a diverse set of addresses with different virtual and physical page numbers, but the same page offset, we should be able to monitor any discrepancy.

\begin{algorithm}[t]
	\begin{algorithmic}
		\caption{Address Aliasing}
		\label{alg:peaks}
		\For {$p$ \textbf{from} w \textbf{to} PAGE\_COUNT} 
			\For {$i$ \textbf{from} w \textbf{to} $0$}
				\State $data\xrightarrow{store}buffer[(p-i)\times PAGE\_SIZE]$
			\EndFor
			\State $t_1 = rdtscp()$
				\State $data\xleftarrow{load}buffer[x\times PAGE\_SIZE]$
			\State $t_2 = rdtscp()$
			\State $measure[p] \gets t_2 - t_1$
		\EndFor
		\State \Return $measure$
	\end{algorithmic}
\end{algorithm}


\subsection{Speculative Dependency Analysis}
\label{sec:hpc}

\begin{figure}[t!]
  \centering
  \subfigure[Step-wise peaks with a very high latency can be observed on some of the virtual pages]{\includegraphics[width=0.48\textwidth]{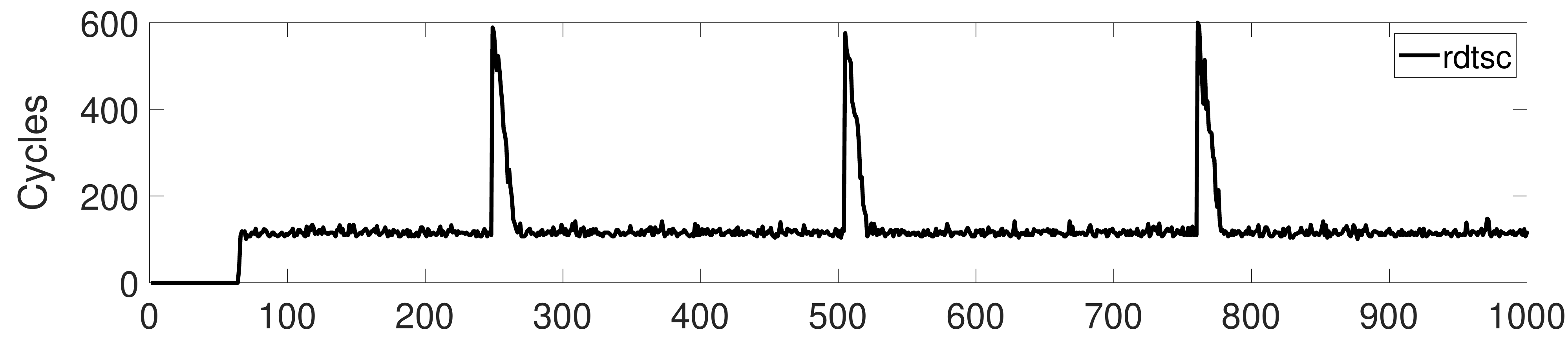}\label{fig:leaka}}
  \subfigure[Affected HPC event:  \texttt{Cycle\_Activity:Stalls\_Ldm\_Pending}]{\includegraphics[width=0.48\textwidth]{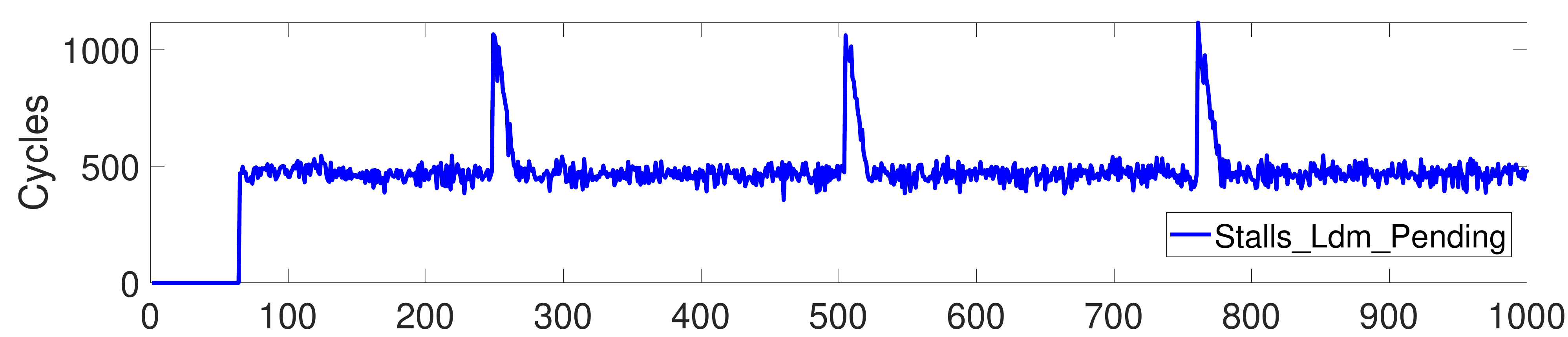}\label{fig:leakb}}
  \subfigure[Affected HPC event: \texttt{Ld\_Blocks\_Partial:Address\_Alias} ]{\includegraphics[width=0.48\textwidth]{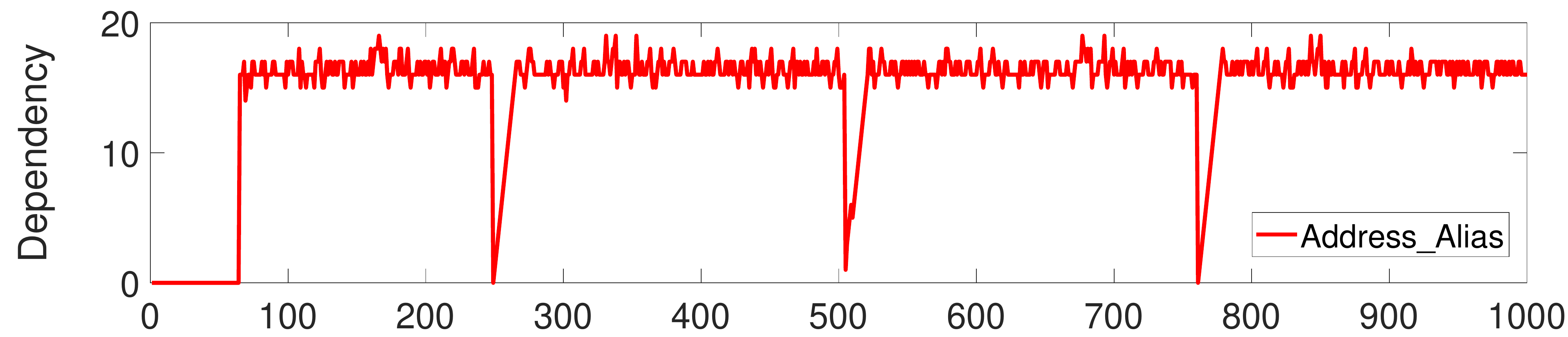}\label{fig:leakc}}
  \caption{\attack's timing measurements and hardware performance counters recorded simultaneously.}
  \label{fig:leak}
\end{figure}

In this section, we use~\autoref{alg:peaks} and Hardware Performance Counters (HPC) to perform an empirical analysis of the dependency resolution logic. HPCs can keep track of low-level hardware-related events in the CPU. The counters are accessible via special purpose registers and can be used to analyze the performance of a program. They provide a powerful tool to detect microarchitectural components that cause bottlenecks. Software libraries such as Performance Application Programming Interface (PAPI)~\cite{terpstra2010collecting} simplifies programming and reading low-level HPC on Intel processors. Initially, we execute~\autoref{alg:peaks} for 1000 different virtual pages. Figure \autoref{fig:leaka} shows the cycle count for each iteration with a set of 4\,kB aliased store addresses. Interestingly, we observe multiple step-wise peaks with a very high latency. Then, we use PAPI to monitor $30$ different performance counters listed in~\autoref{table:fullcounter} in the appendix while running the same experiment. At each iteration, only one performance counter is monitored alongside the aforementioned timing measurement. After each speculative load, the performance counter value and the \texttt{load} time are both recorded. Finally, we obtain the timings and performance counter value pairs as depicted in \autoref{fig:leak}.

  


\begin{figure}[t!]
  \includegraphics[width=1\linewidth]{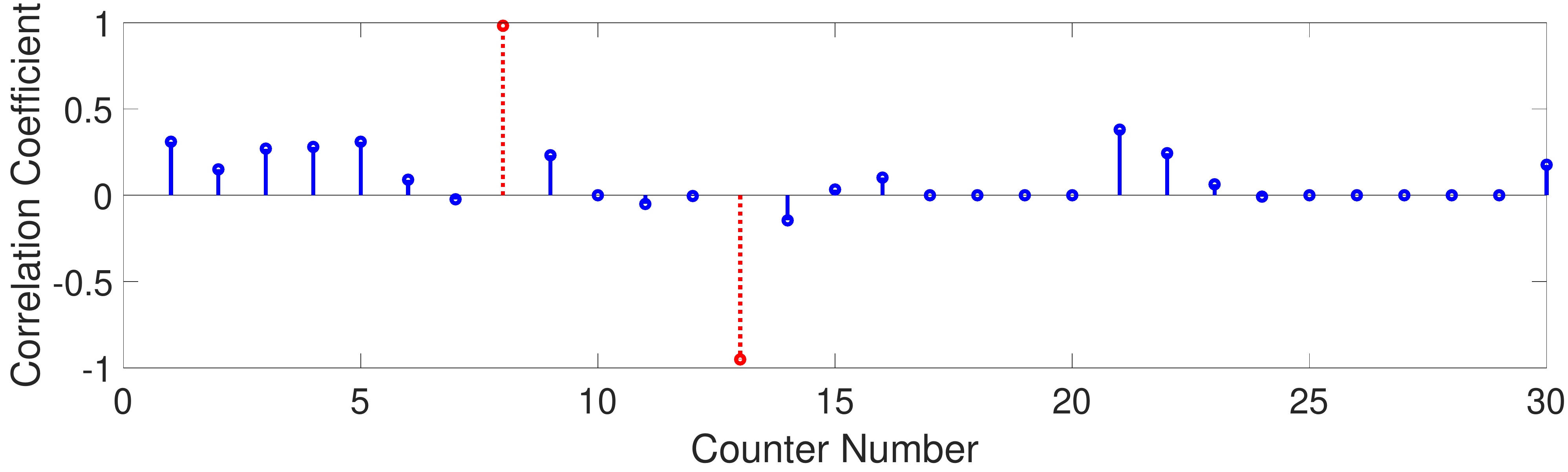}
  \caption{Correlation with HPCs listed in~\autoref{table:fullcounter} in the appendix. \texttt{Ld\_Blocks\_Partial:Address\_Alias} and \texttt{Cycle\_Activity:Stalls\_Ldm\_Pending} (both dotted red) have strong positive and negative correlations, respectively.}
  \label{fig:correlation}
\end{figure}

To find any relation between the observed high latency and a particular event, we compute correlation coefficients between counters and the timing measurements. Since the latency only occurs in the small region of the trace where the timing increases, we only need to compute the correlation on these regions. When an increase of at least 200 clock cycles is detected, the next $s$ values from timing and the HPC traces are used to calculate the correlations, where $s$ is the number of steps from~\autoref{tab:leakyproc} and 200 is the average execution time for a load.

As shown in ~\autoref{fig:correlation}, two events have a high correlation with the leakage: \texttt{\small{Cycle\_Activity:Stalls\_Ldm\_Pending}} has the highest correlation of $0.985$. This event shows the number of cycles for which the execution is stalled and no instructions are executed due to a pending \texttt{load}. \texttt{\small{Ld\_Blocks\_Partial:Address\_Alias}} has an inverse correlation with the leakage. This event counts the number of false dependencies in the MOB when loosenet resolves the 4K aliasing condition. Separately, \texttt{\small{Exe\_Activity:Bound\_on\_Stores}} increases with more number of \texttt{stores} within the inner window loop in \autoref{alg:peaks}, but it does not have a correlation with the leakage. The reason behind this behavior is that the store buffer is full, and additional store operations are pending. However, since there is no correlation with the leakage, this shows that the timing behavior is not due to the \texttt{stores} delay. We also attempt to profile any existing counters related to the memory disambiguation. However, the events \texttt{\small{Memory\_Disambiguation.Success}} and \texttt{\small{Memory\_Disambiguation.Reset}} are not available on the modern architectures that are tested.

\subsection{Leakage of the Physical Address Mapping}

In this experiment, we evaluate whether the observed step-wise latency has any relationship with the physical page numbers by observing the \texttt{pagemap} file. As shown in \autoref{fig:leaksteps}, we observe step-wise peaks with a very high latency which appear once in every 256 pages on average.The 20 least significant bits of physical address for the \texttt{load} matches with the physical addresses of the \texttt{stores} where high peaks for virtual pages are observed. In our experiments, we always detect peaks with different virtual addresses, which have the matching least 20 bits of physical address. This observation clearly discovers the existence of 1\,MB aliasing effect based on the physical addresses. This 1\,MB aliasing leaks information about 8 bits of mapping that were unknown to the user space processes.

Matching this observation with the previously observed \texttt{\small{Cycle\_Activity:Stalls\_Ldm\_Pending}} with a high correlation, the speculative load has been stalled to resolve the dependency with conflicting store buffer entries after the occurrence of a 1\,MB aliased address. This observation verifies that the latency is due to the pending \texttt{load}. When the latency is at the highest point, \texttt{\small{Ld\_Blocks\_Partial:Address\_Alias}} drops to zero, and it increments at each down step of the peak. This implies that the loosenet check does not resolve the rest of the store dependencies whenever there is a 1\,MB aliased address in the store buffer. 

\subsection{Evaluation}

\begin{table}[t!]
\centering
\begin{tabular}{| c | c | c | c | c |} 
 \hline
 \small{CPU Model} & \small{Architecture} & \small{Steps} & \small{SB Size}\\ [0.5ex] 
 \hline\hline
\small{Intel Core i7-8650U} & \small{Kaby Lake R} & \small{22} & \small56\\
\small{Intel Core i7-7700} & \small{Kaby Lake} & \small{22} & \small56\\
\small{Intel Core i5-6440HQ} & \small{Skylake} & \small{22} & \small{56}\\ 
\small{Intel Xeon E5-2640v3} & \small{Haswell} & \small{17} & \small{42} \\
\small{Intel Xeon E5-2670v2} & \small{Ivy Bridge EP} & \small{14} & \small36\\
\small{Intel Core i7-3770} & \small{Ivy Bridge} & \small{12} & \small36\\
\small{Intel Core i7-2670QM} & \small{Sandy Bridge} & \small{12} & \small{36}\\
\small{Intel Core i5-2400} & \small{Sandy Bridge} & \small{12} & \small{36}\\
\small{Intel Core i5 650} & \small{Nehalem} & \small{11} & \small{32}\\
\small{Intel Core2Duo T9400} & \small{Core} & \small{N/A} & \small{20}\\
\small{Qualcomm Kryo 280} & \small{ARMv8-A} & \small{N/A} & \small{*}\\
\small{AMD A6-4455M} & \small{Bulldozer} & N/A & \small{*}\\
 \hline
\end{tabular}
\caption{1\,MB aliasing on various architectures: The tested AMD and ARM architectures, and Intel Core generation do not show similar effects. The Store Buffer (SB) sizes are gathered from Intel Manual~\cite{inteloptimze} and \textit{wikichip.org}~\cite{ivy_wikichip,skylake_wikichip,kabylake_wikichip}.}
\label{tab:leakyproc}
\vspace{-2ex}
\end{table}

\begin{figure*}[t!]
    \centering
    \includegraphics[width=.75\linewidth]{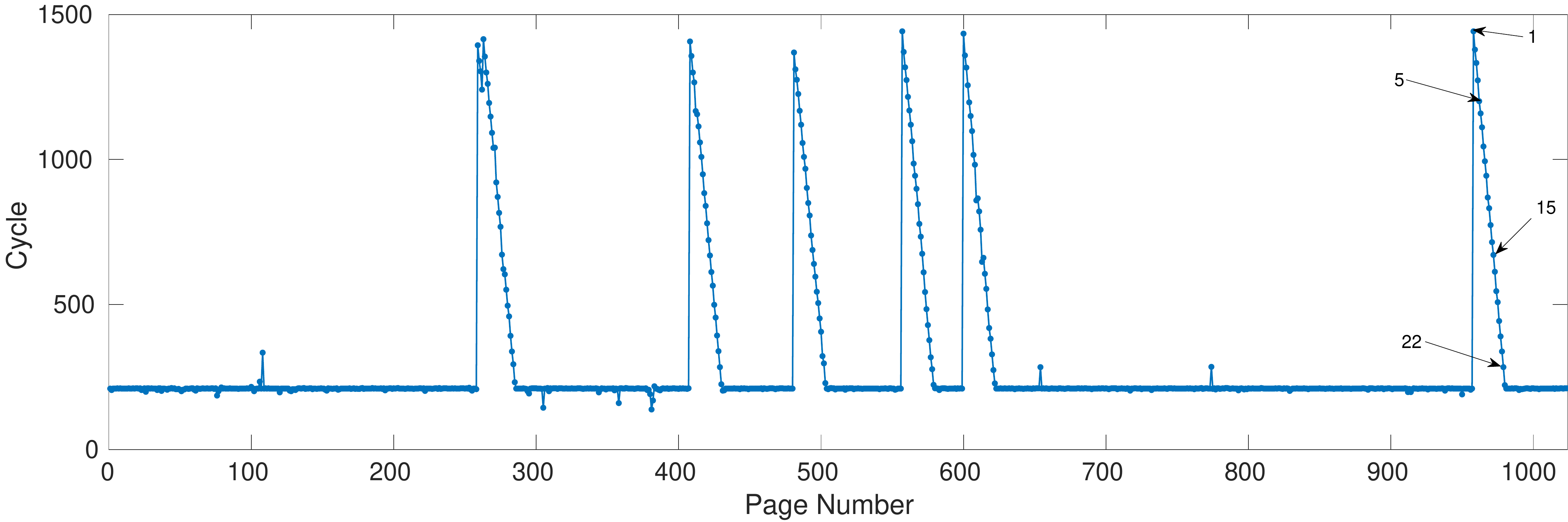}%
    \caption{Step-wise peaks with 22 steps and a high latency can be observed on some of the pages (\textit{Core i7-8650U} processor).}
    \label{fig:leaksteps}
\end{figure*}

In the previous experiment, the execution time of the \texttt{load} operation that is delayed by 1\,MB aliasing decreases gradually in each iteration~(\autoref{fig:leaksteps}). The number of steps to reach the normal execution time is consistent on the same processor. When the first store in the window loop accesses a memory address with the matching 1\,MB aliased address, the latency is at its highest point, marked as ``1'' in \autoref{fig:leaksteps}. As the window loop accesses this address later in the loop, it appears closer to the \texttt{load} with a lower latency like the steps marked as 5, 15 and 22. This observation matches the \textit{carry chain algorithm} described by Intel ~\cite{hily2009resolving} where the aliasing check starts from the most recent \texttt{store}. As shown in \autoref{tab:leakyproc}, experimenting with various processor generations shows that the number of steps has a linear correlation with the size of the store buffer which is architecture dependent. While the leakage exists on all Intel Core processors starting from the first generation, the timing effect is higher for the more recent generations with a bigger store buffer size. The analyzed ARM and AMD processors do not show similar behavior\footnote{We use \texttt{rdtscp} for Intel and AMD processors and the \texttt{clock\_gettime} for ARM processors to perform the time measurements.}.

As our time measurement for speculative load suggests, it is not possible to reason whether the high timing is due to a very slow \texttt{load} or commitment of store operations. If the step-wise delay matches the store buffer entries, this delay may be either due to the the dependency resolution logic performing a pipeline flush and restart of the \texttt{load} for each 4\,kB aliased entry starting from the 1\,MB aliased entry, or due to the \texttt{load} waiting for all the remaining \texttt{stores} to commit because of an unresolved hazard. To explore this further, we perform an additional experiment with all store addresses replaced with non-aliased addresses except for one. This experiment shows that the peak disappears if there is only a single 4\,kB and 1\,MB aliased address in the store buffer.

Lastly, we run the same experiments on a shuffled set of virtual addresses to assure that the contiguous virtual addresses may not affect the observed leakage. Our experiment with the shuffled virtual addresses exactly match the same step-wise behavior suggesting that the upper bits in virtual addresses do not affect the leakage behavior, and the leakage is solely due to the aliasing on physical address bits.

\begin{figure}[t!]
  \includegraphics[width=.95\linewidth]{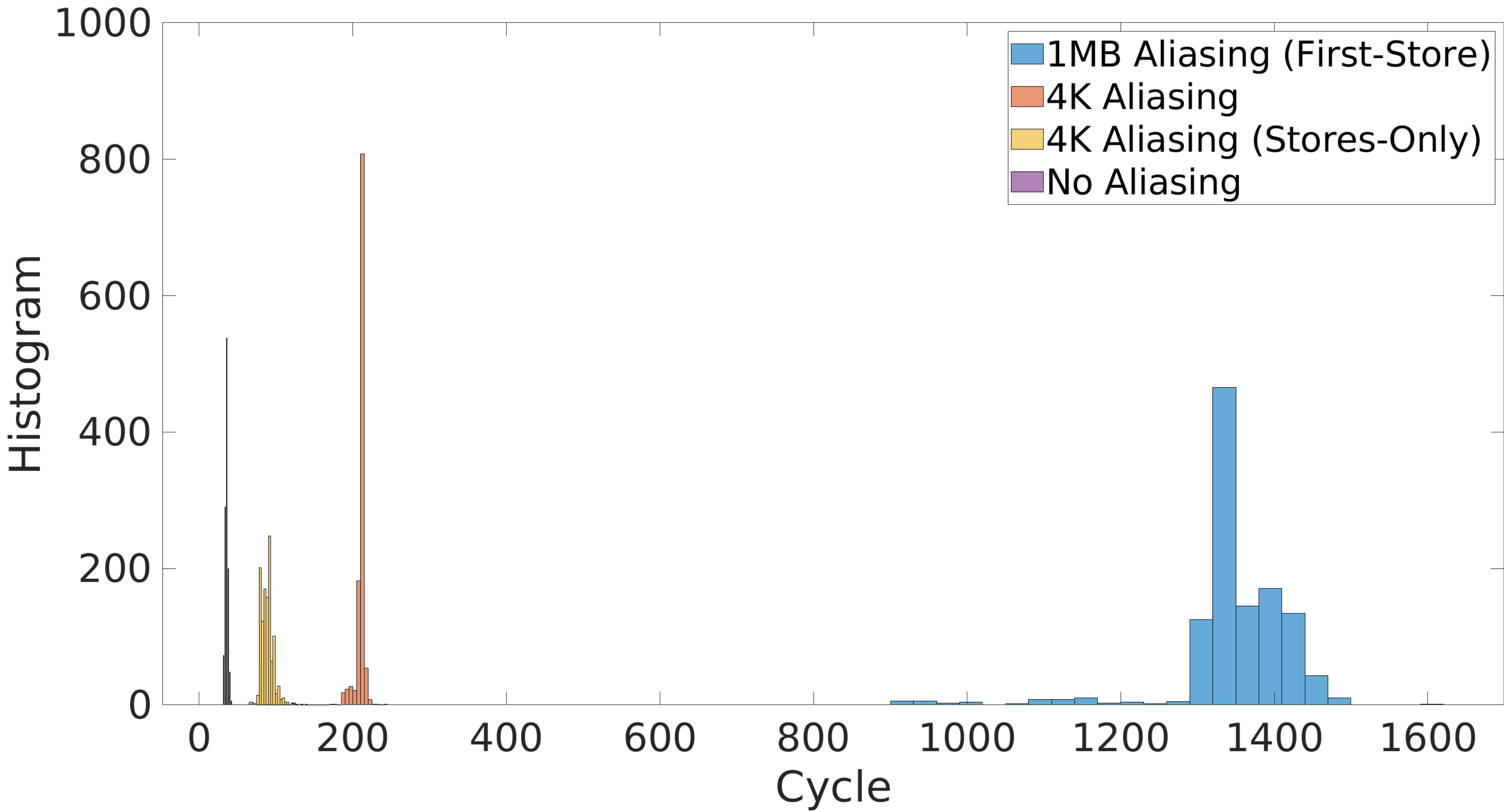}
  \caption{Histogram of the measurement for the speculative load with various store addresses. \texttt{Load} will be fast, $30$ cycles, without any dependency. If there exists 4K aliasing only between the \texttt{stores}, the average is $100$. The average is $200$ when there is 4K aliasing of \texttt{load} and \texttt{stores}. The 1\,MB aliasing has a distinctive high latency.}
  \label{fig:leakhist}
\end{figure}

\subsubsection{Comparison of Address Aliasing Scenarios}

We further test other address combinations to compare additional address aliasing scenarios using~\autoref{alg:peaks}. As shown by ~\autoref{fig:leakhist}, when \texttt{stores} and the \texttt{load} access different cache sets without aliasing, the \texttt{load} is executed in $30$ cycles, which is the typical timing for an L1 data cache load including the \texttt{rdtscp} overhead. When the \texttt{stores} have different memory addresses with the same page offset, but the \texttt{load} has a different offset, the \texttt{load} takes 100 cycles to execute. This shows that even memory addresses in the store buffer having 4K Aliasing conditions with each other that are totally unrelated to the speculative load create a memory bottleneck for the \texttt{load}. In the next scenario, 4K aliasing between the \texttt{load} and all \texttt{stores}, the average load time is about $200$ cycles. While the aforementioned 4K aliasing scenarios may leak cross domain information about memory accesses (~\autoref{sec:track}), the most interesting scenario is the 1\,MB aliasing which takes more than $1200$ cycles for the highest point in the peak. For simplicity, we refer to the 1\,MB aliased address as \textit{aliased address}, in the rest of the paper.

\subsection{Discussion}
\subsubsection{The Curious Case of Memory Disambiguation}

The processor uses an additional speculative engine, called the \textit{memory disambiguator}~\cite{doweck2006inside,krimer2013counter}, to predict memory false dependencies and reduce the chance of their occurrences. The main idea is to predict if a \texttt{load} is independent of preceding \texttt{stores} and proceed with the execution of the \texttt{load} by ignoring the store buffer. The predictor uses a hash table that is indexed with the address of the \texttt{load}, and each entry of the hash table has a saturating counter. If the pre-commitment dependency resolution does not detect false dependencies, the counter is incremented, otherwise it will be reset to zero. After multiple successful executions of the same \texttt{load} instruction, the predictor assumes that the \texttt{load} is safe to execute. Every time the counter resets to zero, the next iteration of the \texttt{load} will be blocked to be checked against the store buffer entries. Mispredictions result in performance overhead due to pipeline flushes. To avoid repeated mispredictions, a watchdog mechanism monitors the success rate of the prediction, and it can temporarily disable the memory disambiguator. 

The predictor of the memory disambiguator should go into a stable state after the first few iterations, since the memory \texttt{load} is always truly independent of any aliased store. Hence the saturating counter for the target speculative load address passes the threshold, and it never resets due to a false prediction. As a result, the memory disambiguator should always fetch the data into the cache without any access to the store buffer. However, since the memory disambiguation performs speculation, the dependency resolution at some point verifies the prediction. The misprediction watchdog is also supposed to only disable the memory disambiguator when the misprediction rate is high, but in this case we should have a high prediction rate. Accordingly, the observed leakage occurs after the disambiguation and during the last stages of dependency resolution, i.e., the memory disambiguator only performs prediction on the 4K aliasing at the initial loosenet check, and it cannot protect the pipeline from 1\,MB aliasing that appears at a later stage.

\subsubsection{Hyperthreading Effect}
Similar to the 4K Aliasing~\cite{moghimi2018memjam,sullivan2018microarchitectural}, we empirically test whether the 1\,MB aliasing can be used as a covert/side channel through logical processors. Our observation shows that when we run our experiments on two logical processors on the same physical core, the number of steps in the peaks is exactly halved. This matches the description by Intel~\cite{inteloptimze} where it is stated that the store buffer is split between the logical processors. As a result, the 1\,MB aliasing effect is not visible and exploitable across logical cores. ~\cite{kosinski2016store} suggests that loosenet checks mask out the \texttt{stores} on the opposite thread.

\section{\attack\ from JavaScript}\label{sec:Javascript}

\begin{figure}[tp]
  \includegraphics[width=.99\linewidth]{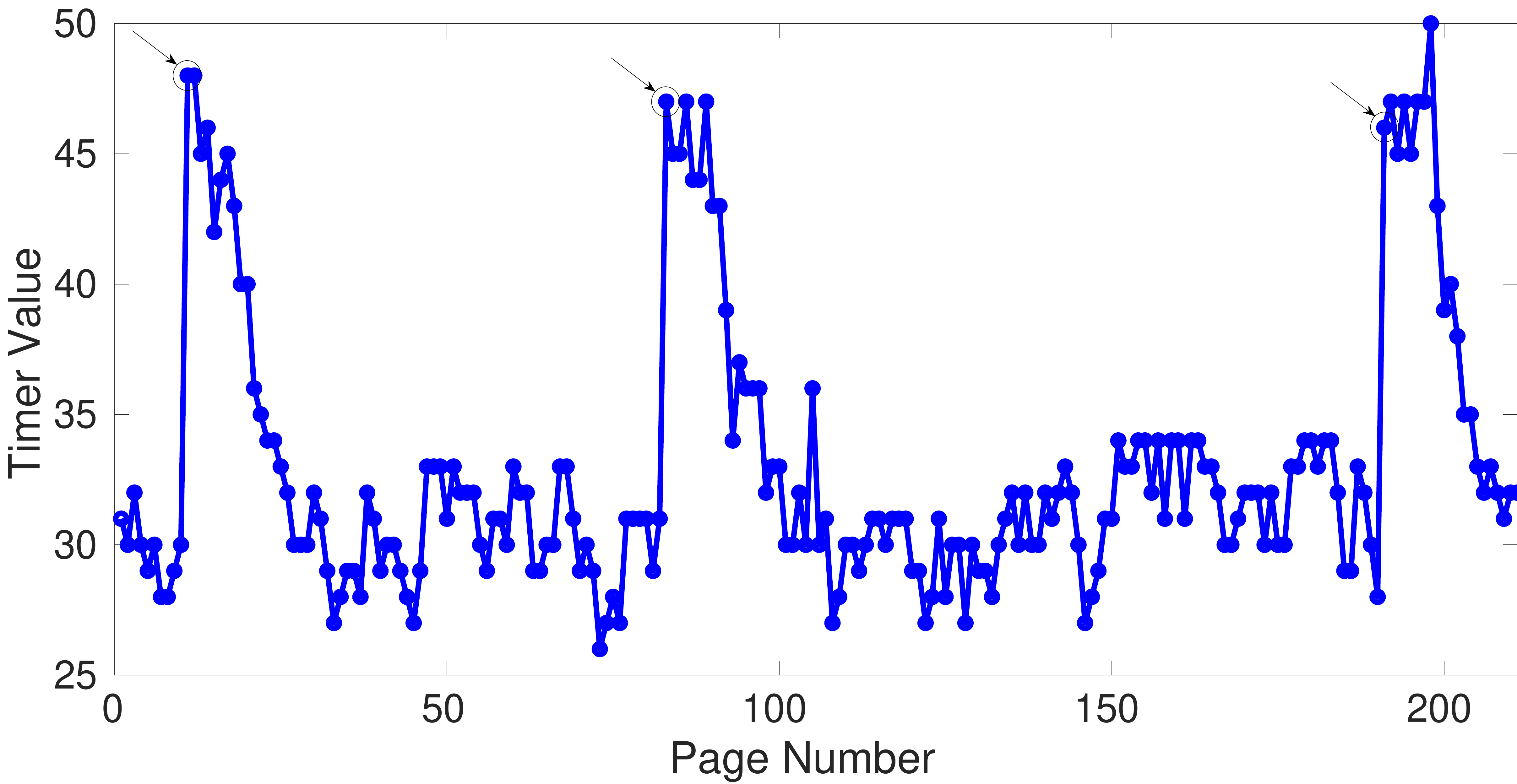}
  \caption{Reverse engineering physical page mappings in JavaScript. The markers point to addresses having same 20 bits of physical addresses being part of the same eviction set.}
  \label{fig:js}
\end{figure}

Microarchitectural attacks from JavaScript have a high impact as drive-by attacks in the browser can be accomplished without any privilege or physical proximity. In such attacks, co-location is automatically granted by the fact that the browser loads a website with malicious embedded JavaScript code. The browsers provide a sandbox where some instructions like \texttt{clflush} and \texttt{prefetch} and file systems such as \texttt{procfs} are inaccessible, limiting the opportunity for attack. Genkin et al.~\cite{genkin2018drive} showed that side-channel attacks inside a browser can be performed more efficiently and with greater portability through the use of \textit{WebAssembly}.
Yet, WebAssembly introduces an additional abstraction layer, i.e. it emulates a 32-bit environment that translates the internal addresses to virtual addresses of the host process (the browser). WebAssembly only uses addresses of the emulated environment and similar to JavaScript, it does not have direct access to the virtual addresses. Using \attack\ from JavaScript opens the opportunity to puncture these abstraction layers and to obtain physical address information directly. \autoref{fig:js} shows the address search in JavaScript using \attack. Compared to native implementations, we replace the \texttt{rdtscp} measurement with a timer based on a shared array buffer \cite{hansen2016shared}. We cannot use any fence instruction such as \texttt{lfence}, and as a result, there remains some negligible noise in the JavaScript implementation. However, the aliased addresses can still be clearly seen, and we can use this information to improve the state-of-the art eviction set creation for both Rowhammer and cache attacks.


\subsection{Efficient Eviction Set Finding}
\label{subsection:eviction_Set}

We use the algorithm proposed in~\cite{genkin2018drive}. It is a slight improvement to the former state-of-the-art brute force method\cite{orenspy} and consists of three phases: 
\begin{itemize}
    \item \textit{expand}: A large pool of addresses $P$ is allocated with the last twelve bits of all addresses being zero. A random address is picked as a witness $t$ and tested against a candidate set $C$. If $t$ is not evicted by $C$, it is added to $C$ and a new witness will be picked. As soon as $t$ gets evicted by $C$, $C$ forms an eviction set for $t$.
    \item \textit{contract}: Addresses are subsequently removed from the eviction set. If the set still evicts $t$, the next address is removed. If it does not evict $t$ anymore, the removed address is added back to the eviction set. At the end of this phase, we have a minimal eviction set of the size of the set associativity. 
    \item \textit{collect}: All addresses mapping to the already found eviction set are removed from $P$ by testing if they are evicted by the found set. After finding 128 initial cache sets, this approach utilizes the linearity property of the cache: For each found eviction set, the bits 6-11 are enumerated instead. This provides 63 more eviction sets for each found set, leading to full cache coverage.
\end{itemize}

We test this approach on an Intel Core i7-4770 with four physical cores and a shared 8MB 16-way L3 cache with Chromium 68.0.3440.106, Firefox 62 and Firefox Developer Edition 63. The approach yields an $80\%$ accuracy rate to find all $8192$ eviction sets when starting with a pool of $4096$ pages. The entire eviction set creation process takes an average of $46\,s$. We improve the algorithm by \textbf{1)} using the addresses removed from the eviction set in the contract phase as a new candidate set and \textbf{2)} removing more than one address at a time from the eviction set during the contract phase. The improved eviction set creation process takes $35\,s$ on average. 

\subsubsection{Evaluation}
The probability of finding a congruent address is $P(C) = 2^{\gamma - c - s}$, where $c$ is the number of bits determining the cache set, $\gamma$ is the number of bits attackers know, and $s$ is the number of slices\cite{vila2018theory}. Since \attack\ allows us to control $\gamma \geq c$ bits, we are only left with uncertainty about a few address bits that influence the slice selection algorithm~\cite{irazoqui2015systematic}. In theory, the eviction set search is sped up by a factor of 4096 by using aliased addresses in the pool, since on average one of $ 2 ^ 8 $ instead of one of $ 2 ^ {20} $ addresses is an aliased address. Additionally, the address pool is much smaller, where $115$ addresses are enough to find all the eviction sets. In native code, the overhead involved in finding the aliased addresses is negligible, less than a second in our experiments. However, in JavaScript, due to the noise, it takes 9s for finding aliased addresses and then 3s for eviction set as compared to the baseline of 46s for classic method in \autoref{tbl:performanceStoreForESSearch}. Success rate however is 100\% with \attack\ as compared to 80\% for the classic method. Besides, success rate of the classical method can be affected by the availability and consumption of memory on the system.

From each aliased address pool, 4 eviction sets can be found (corresponding to the 4 slices which are the only unknown part in the mapping). These can be enumerated again to form 63 more eviction sets since we still kept the bits 6-11 fixed. To accomplish full cache coverage, the aliased address pool has to be constructed 32 times. The \attack\ variant for finding eviction sets is more susceptible to system noise, which is why it needs more repetitions i.e. $R$ rounds to get reliable values. On the other hand, it is less prone to values deviating largely from the mean, which is a problem in the classic eviction set creation algorithm. The classic method does not succeed about one out of five times in our experiments, as shown in \autoref{tbl:performanceStoreForESSearch}. The unsuccessful attempts occur due to aborts if the algorithm takes much longer than statistically expected. As a result, \attack\ can be incorporated in an end-to-end attack such as drive-by key-extraction cache attacks by Genkin et al.~\cite{genkin2018drive}. \attack\ increases both speed and reliability of the eviction set finding and therefore the entire attack.

\begin{table}[t!]
\centering
\begin{tabular}{| c | c | c | c | c | c |}
\hline
Algorithm & $R$ & \begin{tabular}[c]{@{}c@{}}$t_{total}$\end{tabular} & \begin{tabular}[c]{@{}c@{}}$t_{AAS}$\end{tabular} & \begin{tabular}[c]{@{}c@{}}$t_{ESS}$\end{tabular} & \begin{tabular}[c]{@{}c@{}}Success\end{tabular}\\
 \hline  \hline
 Classic\cite{orenspy} & 3 & 46s & - & 100\% & 80\% \\
 Improved~\cite{genkin2018drive} & 3 & 35s & - & 100\% & 80\% \\
 AA (ours) & 10 & 10s & 54\% & 46\% & 67\%\\
 AA (ours) & 20 & 12s & 75\% & 25\% & 100\%                 
\\ 
 \hline
\end{tabular}
\caption{Comparison of different eviction set finding algorithms on an Intel Core i7-4770. Classic is the method from \cite{orenspy}, improved is the same method with slight improvement, \textit{Aliased Address (AA)} uses \attack. $t_{AAS}$ is the time percentage used for finding aliased addresses. $t_{ESS}$ is the time percentage for finding eviction sets. R is the number of Rounds.}
\label{tbl:performanceStoreForESSearch}
\end{table}


\section{Rowhammer Attack using \attack}

To perform a Rowhammer attack, the adversary needs to efficiently access DRAM rows adjacent to a victim row. In a single-sided Rowhammer attack, only one row is activated repeatedly to induce bit flips on one of the nearby rows. For this purpose, the attacker needs to make sure that multiple virtual pages co-locate on the same bank. The probability of co-locating on the same bank is low without the knowledge of physical addresses and their mapping to memory banks. In a double-sided Rowhammer attack, the attacker tries to access two different rows $n+1$ an $n-1$ to induce bit flips in the row $n$ placed between them. While double-sided Rowhammer attacks induce bit flips faster due to the extra charge on the nearby cells of the victim row $n$, they further require access to contiguous memory pages. In this section, we show that \attack\ can help boosting both single and double-sided Rowhammer attacks by its additional 8-bit physical address information and resulting detection of contiguous memory. 

\subsection{DRAM Bank Co-location}
\textit{DRAMA}~\cite{pessl2016drama} reverse engineered the memory controller mapping. This requires elevated privileges to access physical addresses from the \texttt{pagemap} file. The authors have suggested that prefetch side-channel attacks~\cite{gruss2016prefetch} may be used to gain physical address information instead. \attack\ is an alternative way to obtain partial address information and is still feasible when the {\tt prefetch} instruction is not available, e.g. in JavaScript. In our approach, we use \attack\ to detect aliased virtual memory addresses where the 20 LSBs of the physical addresses match. The memory controller uses these bits for mapping the physical addresses to the DRAM banks~\cite{pessl2016drama}. Even though the memory controller may use additional bits, the majority of the bits are known using \attack. An attacker can directly hammer such aliased addresses to perform a more efficient single-sided Rowhammer attack with a significantly increased probability of hitting the same bank. As shown in Table \ref{table:drama}, we reverse engineer the DRAM mappings for different hardware configurations using the DRAMA tool, and only a few bits of physical address entropy beyond the 20 bits will remain unknown.

\begin{table}[t]
\centering
\begin{tabular}{|c | c | c| c |} 
 \hline
 System Model & DRAM Configuration & \# of Bits \\ [0.5ex] 
 \hline\hline
 Dell XPS-L702x & 1 x (4GB 2Rx8) & 21 \\
 (Sandy Bridge) & 2 x (4GB 2Rx8) & 22 \\\hline
 Dell Inspiron-580 & 1 x (2GB 2Rx8) \textbf{(b)} & 21 \\
  (Nehalem) & 2 x (2GB 2Rx8) \textbf{(c)} & 22 \\
 & 4 x (2GB 2Rx8) \textbf{(d)}  & 23 \\\hline
 Dell Optiplex-7010 & 1 x (2GB 1Rx8) \textbf{(a)} & 19 \\
 (Ivy Bridge) & 2 x (2GB 1Rx8) & 20 \\
 & 1 x (4GB 2Rx8) \textbf{(e)} & 21 \\
 & 2 x (4GB 2Rx8) & 22 \\
 \hline
\end{tabular}

\caption{Reverse engineering the DRAM memory mappings using DRAMA tool, \textit{\# of Bits} represents the number of physical address bits used for the bank, rank and channel~\cite{pessl2016drama}.}
\label{table:drama}
\end{table}

\begin{figure}[t!]
  \centering
  \subfigure[19 bits used by memory controller, no unknown bits]{\includegraphics[width=0.47\textwidth]{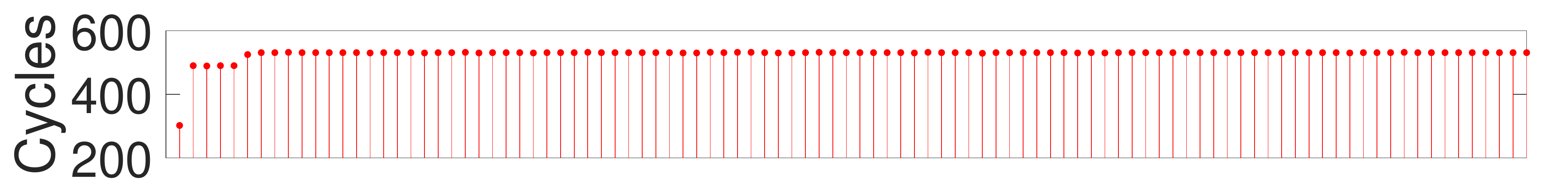}\label{fig:probabilities1}}
  \subfigure[21 bits used by memory controller, 1 unknown bit]{\includegraphics[width=0.47\textwidth]{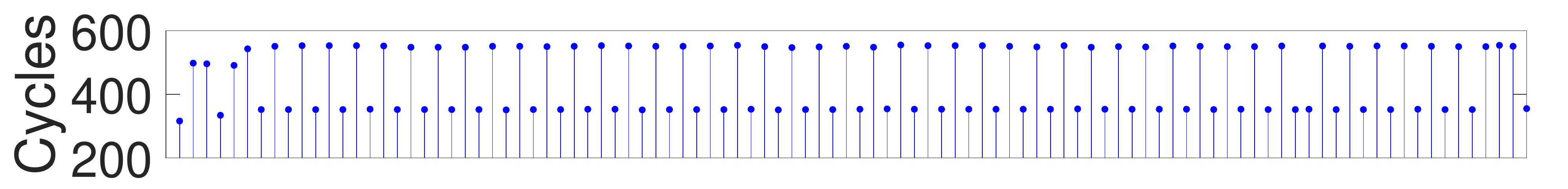}\label{fig:probabilities2}}
  \subfigure[22 bits used by memory controller, 2 unknown bits]{\includegraphics[width=0.47\textwidth]{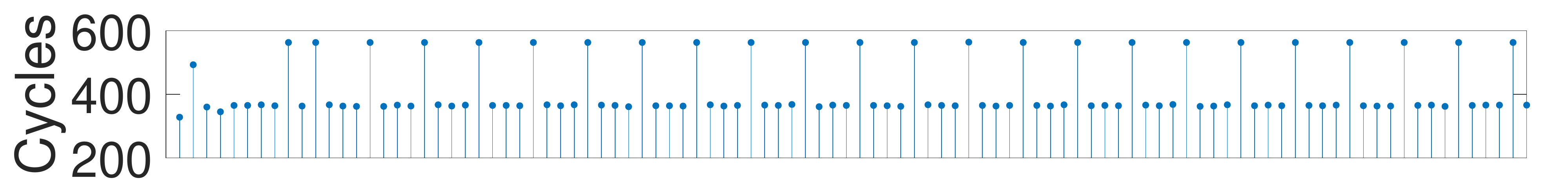}\label{fig:probabilities3}}
  \subfigure[23 bits used by memory controller, 3 unknown bits]{\includegraphics[width=0.47\textwidth]{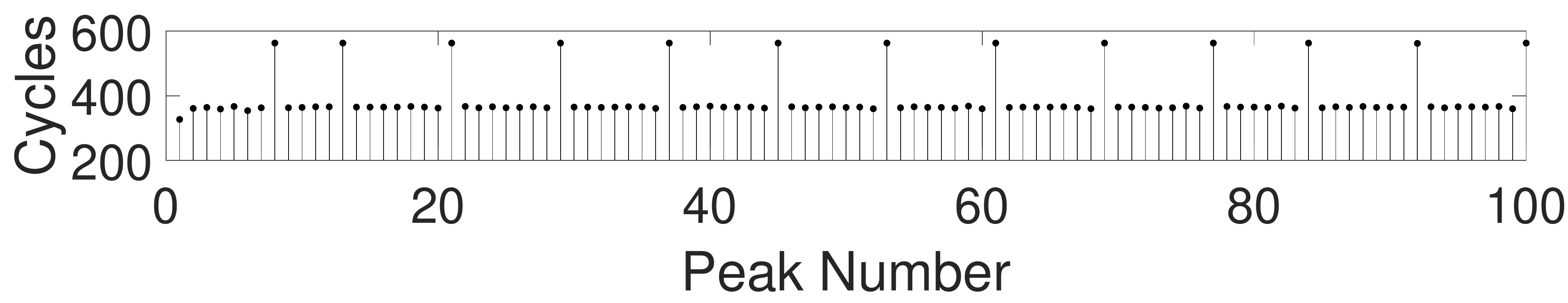}\label{fig:probabilities4}}
  \caption{Bank co-location for various DRAM configurations (a), (b), (c) \& (d) from~\autoref{table:drama}. The regularity of the peaks shows that the allocated memory was contiguous, which is coincidental.}
  \label{fig:probabilities}
\end{figure}

To verify if our aliased virtual addresses co-locate on the same bank, we use the row conflict side channel as proposed in~\cite{frigo2018grand} (timings in the appendix, \autoref{subsec:row-conflict}). We observe that whenever the number of physical address bits used by the memory controller to map data to physical memory is equal to or less than 20, we always hit the same bank. For each additional bit the memory controller uses, the probability of hitting the same bank is divided by 2 as there is one more bit of entropy. In general, we can formulate that our probability $p$ to hit the same bank is $p = 1/2^n$, where $n$ is the number of unknown physical address bits in the mapping. We experimentally verify the success rate for the setups listed in~\autoref{table:drama}, as depicted in~\autoref{fig:probabilities}. In summary, \attack\ drastically improves the efficiency of finding addresses mapping to the same bank without administrative privilege or reverse engineering the memory controller mapping.


\subsection{Contiguous Memory}
\begin{figure}[tp]
  \includegraphics[width=1\linewidth]{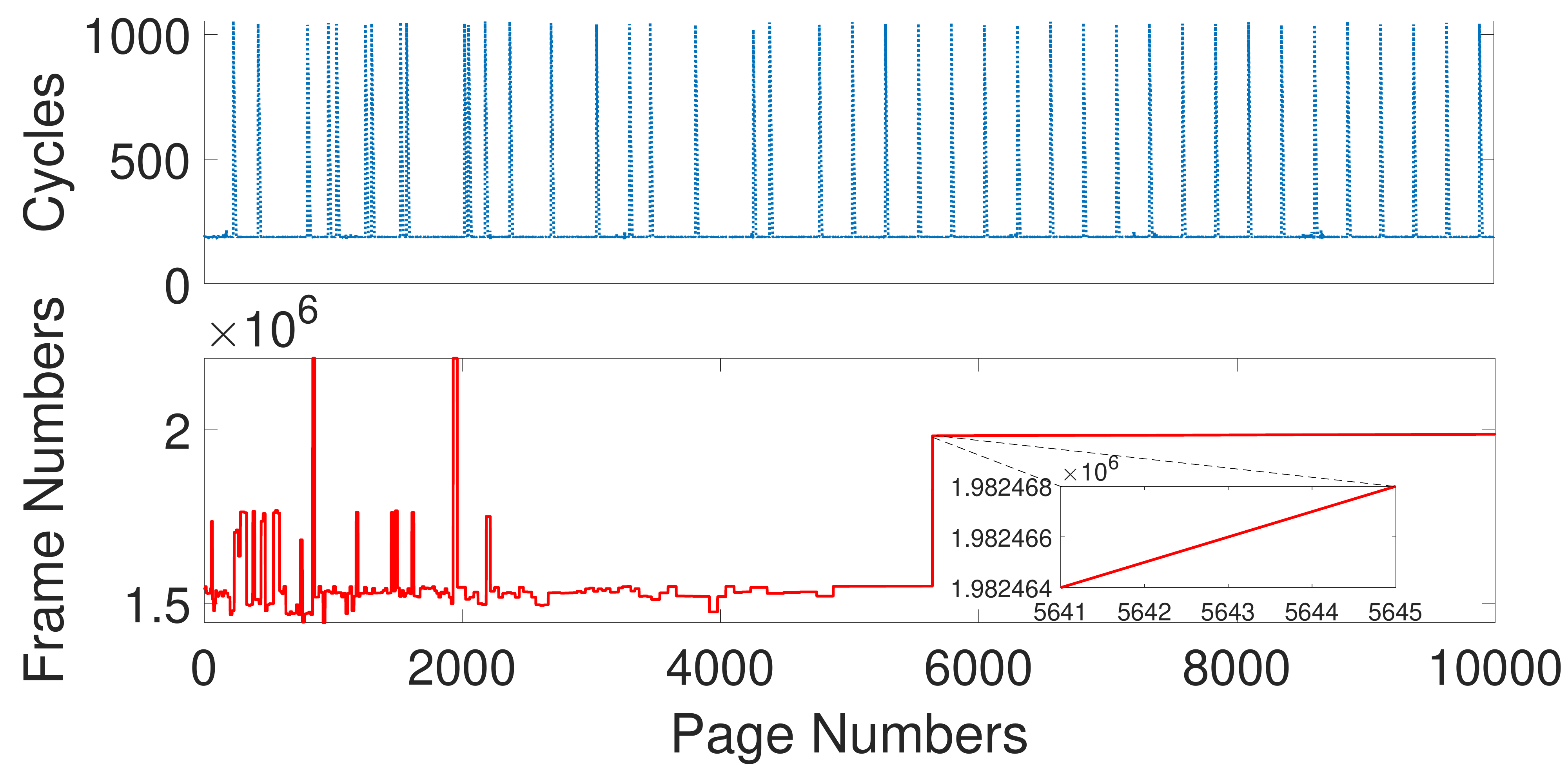}
  \caption{Relation between leakage peaks and the physical page numbers. The dotted plot shows the leakage peaks from \attack. The solid plot shows the decimal values of the physical frame numbers from the {\tt pagemap} file. Once the peaks in the dotted plot become regular, the solid plot is linearly increasing, which shows contiguous memory allocation.}
  \label{fig:cont_mem3}
\end{figure}

For a double-sided Rowhammer attack, we need to hammer rows adjacent to the victim row in the same bank. This requires detecting contiguous memory pages in the allocated memory, since the rows are written to the banks sequentially.  Without contiguous memory, the banks will be filled randomly and we will not be able to locate neighboring rows. We show that an attacker can use \attack\ to detect contiguous memory using 1 MB\, aliasing peaks. For this purpose, we compare the physical frame numbers to the \attack\ leakage for $10000$ different virtual pages allocated using \texttt{malloc}. ~\autoref{fig:cont_mem3} shows the relation between 1 MB\, aliasing peaks and physical page frame numbers. When the distance between the peaks is random, the trend of frame numbers also change randomly. After around $5000$ pages, we observe that the frame numbers increase sequentially. The number of pages between the peaks remains constant at $256$ where this distance comes from the 8 bits of physical address leakage due to 1 MB\, aliasing. 

We also compare the accuracy of obtaining contiguous memory detected by \attack\, by analyzing the actual physical addresses from the {\tt pagemap} file. By checking the difference between physical page numbers for each detected virtual page, we can determine the accuracy of our detection method: the success rate for finding contiguous memory is above 99\% disregarding the availability of the contiguous pages. For detailed experiment on the availability of the contiguous pages, see \autoref{sec:memutil} in the appendix.

\subsection{Double-Sided Rowhammer with \attack}
\label{subsec:row_spoiler}
As double-sided Rowhammer attacks are based on the assumption that rows within a bank are contiguous, we mount a practical double-sided Rowhammer attack on several DRAM modules using \attack\ without any root privileges. First, we use \attack\ to detect a suitable amount of contiguous memory. If enough contiguous memory is available in the system, \attack\ finds it, otherwise a double-sided Rowhammer attack is not feasible. In our experiments, we empirically configure \attack\ to detect 10\,MB of contiguous memory. Second, we apply the row conflict side channel only to the located contiguous memory, and get a list of virtual addresses which are contiguously mapped within a bank. Finally, we start performing a double-sided Rowhammer attack by selecting 3 consecutive addresses from our list. While we have demonstrated the bit flips in our own process, we can free that memory which can then be assigned to a victim process by using previously known techniques like spraying and memory waylaying \cite{gruss2018another}. As the bit flips are highly reproducible, we can again flip the same bits in the victim process to demonstrate a full attack. \autoref{tab:flippydrams} shows some of the DRAM modules susceptible to Rowhammer attack.

The native version of Rowhammer in this work is also applicable in JavaScript. The JavaScript-only variant implementation of Rowhammer by Gruss et al. \cite{gruss2016rowhammer}, named rowhammer.js\footnote{\url{https://github.com/IAIK/rowhammerjs}}, can be combined with \attack\ to implement an end-to-end attack. In the original rowhammer.js, 2MB huge pages were assumed to get a contiguous chunk of physical memory. With \attack, this assumption is no longer required as explained in \autoref{subsec:row_spoiler}.



\begin{table}[t!]
\centering
\begin{tabular}{|c | c | c |} 
 \hline
 DRAM Model & Architecture & Flippy \\ [0.5ex] 
 \hline\hline
\small{M378B5273DH0-CK0} & Ivy Bridge & \checkmark \\
\small{M378B5273DH0-CK0} & Sandy Bridge & \checkmark \\
\small{M378B5773DH0-CH9} & Sandy Bridge & \checkmark \\ 
\small{M378B5173EB0-CK0} & Sandy Bridge & $\times$ \\
\small{NT2GC64B88G0NF-CG} & Sandy Bridge & $\times$ \\
\small{KY996D-ELD} & Sandy Bridge & $\times$ \\
\small{M378B5773DH0-CH9} & Nehalem & \checkmark \\
\small{NT4GC64B8HG0NS-CG} & Sandy Bridge & $\times$ \\
\small{HMA41GS6AFR8N-TF} & Skylake & $\times$  \\
 \hline
\end{tabular}
\caption{DRAM modules susceptible to double-sided Rowhammer attack using \attack.}
\label{tab:flippydrams}
\end{table}

\begin{figure}[t!]
  \includegraphics[width=.95\linewidth]{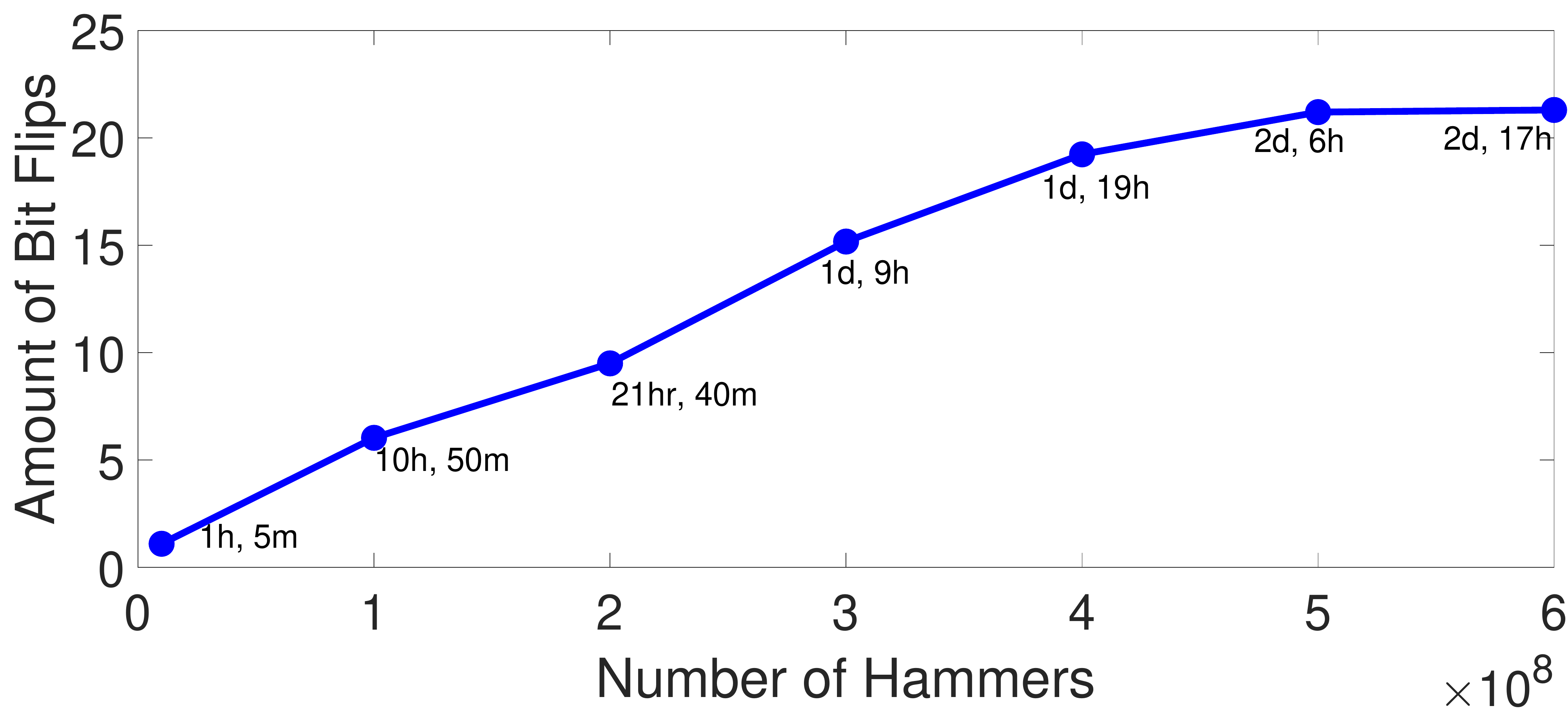}
  \caption{Amount of bit flips increases with the increase in number of hammerings.
  The timings do not include the time taken for reboots and 1 minute sleep time.}
  \label{fig:bitflips}
\end{figure}

\autoref{fig:bitflips} shows the number of hammers compared to the the amount of bit flips for configuration (e) in \autoref{table:drama}. We repeat this experiment 30 times for every measurement and the results are then averaged out. On every experiment, the system is rebooted using a script because once the memory becomes fragmented, no more contiguous memory is available. The number of bit flips increases with more number of hammerings. Hammering for 500 million times is found to be an optimal number for this DRAM configuration, as the continuation of hammering is not increasing bit flips.

\section{Tracking Speculative Loads With \attack} \label{sec:track}
Single-threaded attacks can be used to steal information from other security contexts running before/after the attacker code on the same thread~\cite{SGXpectre18,moghimi2017cachezoom}. Example scenarios are I) context switches between processes of different users, or II) between a user process and a kernel thread, and III) Intel Software Guard eXtensions (SGX) secure enclaves~\cite{moghimi2017cachezoom,van2018nemesis}. In such attacks, the adversary puts the microarchitecture to a particular state, waits for the context switch and execution of the victim thread, and then tries to observe the microarchitectural state after the victim's execution. We propose an attack where the adversary \textbf{1)} fills the store buffer with arbitrary addresses, \textbf{2)} issues the victim context switch and lets the victim perform a secret-dependent memory access, and \textbf{3)} measures the execution time of the victim. Any correlation between the victim's timing and the load address can leak secrets~\cite{yarom2017cachebleed}. Due to the nature of \attack, the victim should access the memory while there are aliased addresses in the store buffer, i.e. if the \texttt{stores} are committed before the victim's speculative load, there will be no dependency resolution hazard. 

\begin{figure}[tp]
  \includegraphics[width=0.48\textwidth]{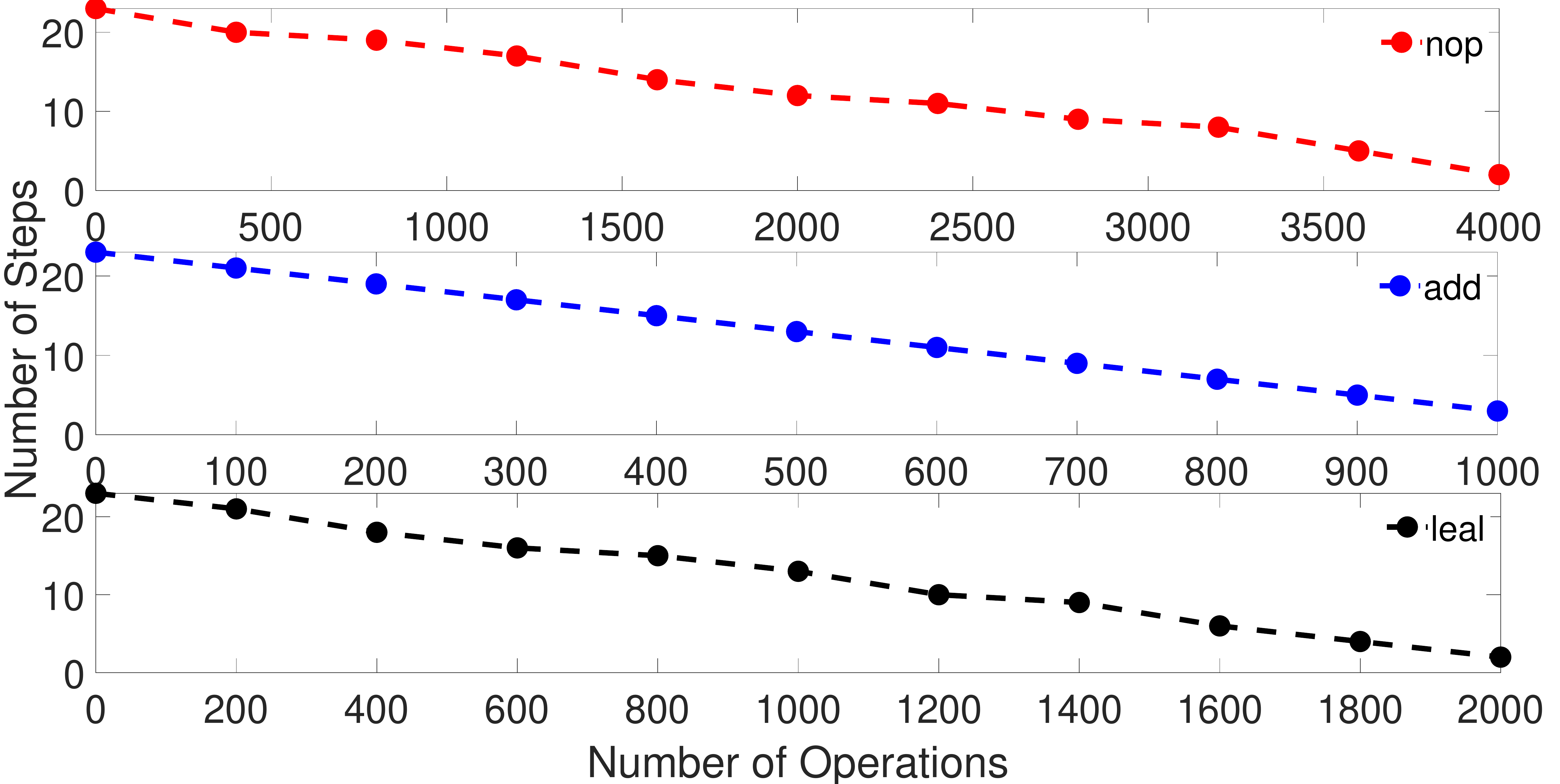}
  \caption{The depth of \attack\ leakage with respect to different instructions and execution units.}
  \label{fig:specdepth}
\end{figure}

We first perform an analysis of the depth of the operations that can be executed between the \texttt{stores} and the \texttt{load} to investigate the viability of \attack. In this experiment, we repeat a number of instructions between \texttt{stores} and the \texttt{load} that are free from memory operations. ~\autoref{fig:specdepth} shows the number of stall steps due to the dependency hazard with the added instructions. Although \texttt{nop} is not supposed to take any cycle, adding 4000 \texttt{nop} will diffuse the timing latency. Then, we test \texttt{add} and \texttt{leal}, which use the Arithmetic Logic Unit (ALU) and the Address Generation Unit (AGU), respectively. ~\autoref{fig:specdepth} shows that only $1000$ \texttt{adds} can be executed between the \texttt{stores} and \texttt{load} before the \attack\ effect is lost. Since each \texttt{add} typically takes about $1$ cycle to execute, this roughly gives a $1000$ cycle depth for \attack. Considering the observed depth, we discuss potential attacks that can track the speculative load in the following two scenarios.
\subsection{\attack\ Context Switch} \label{sec:context-switch}
\begin{figure}[tp]
  \includegraphics[width=.98\linewidth]{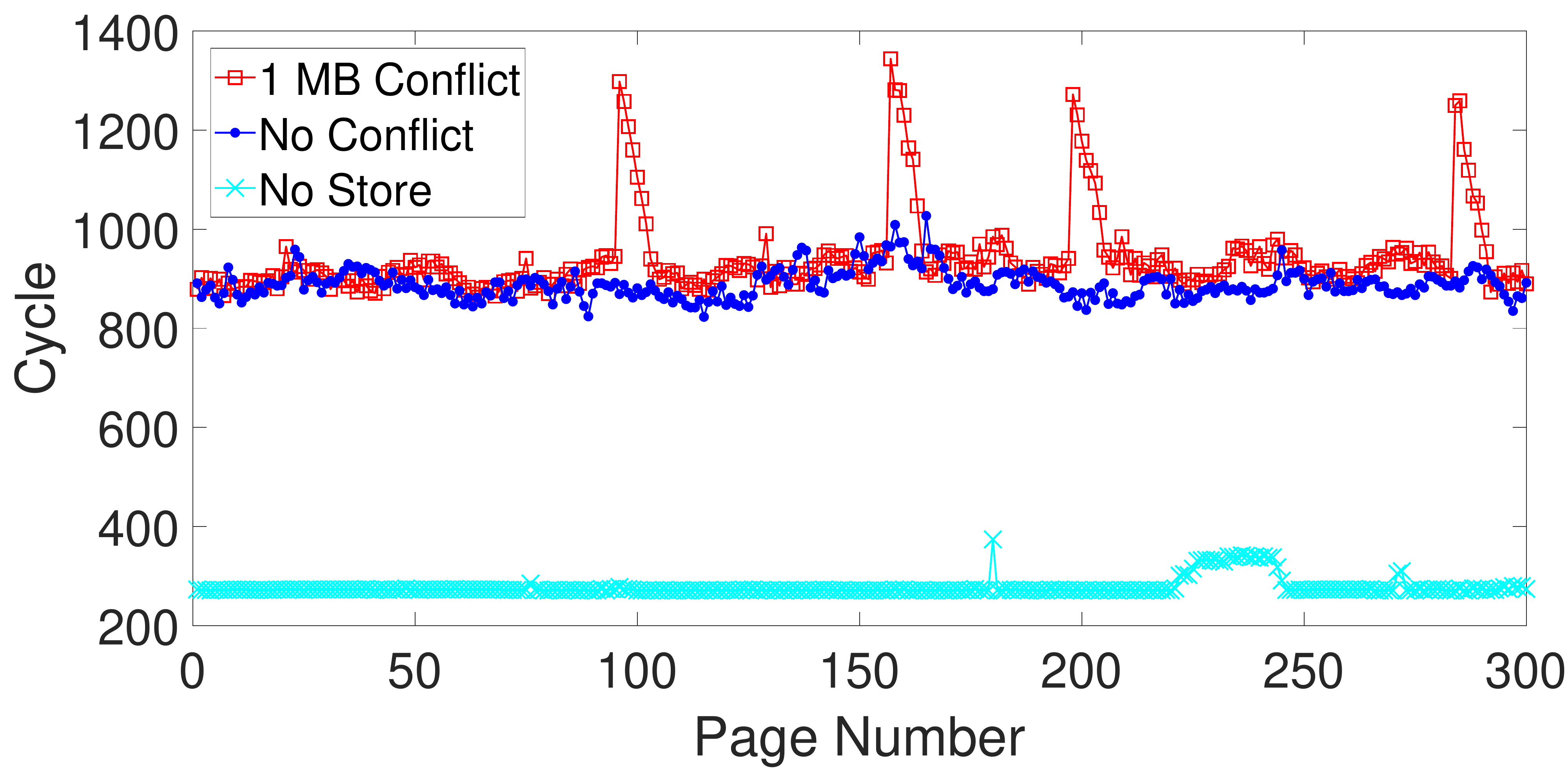}
  \caption{Execution time of \texttt{mincore} system call. When a kernel \texttt{load} address has aliasing with the attacker's \texttt{stores} (red/1MB Conflict), the step-wise delay will appear. These timings are measured with Kernel Page Table Isolation disabled.}
  \label{fig:mincore}
\end{figure}

In this attack, we are interested in tracking a memory access in the privileged kernel environment after a context switch. First, we fill the store buffer with addresses that have the same page offset, and then execute a system call. During the execution of the system call, we expect to observe a delayed execution if a secret \texttt{load} address has aliasing with the \texttt{stores}. We utilize \attack\ to iterate over various virtual pages, thus some of the pages have more noticeable latency due to the 1\,MB aliasing. We analyze multiple \texttt{syscalls} with various execution times. For instance,~\autoref{fig:mincore} shows the execution time for \texttt{mincore}. In the first experiment (red/1 MB Conflict), we fill the store buffer with addresses that have aliasing with a memory \texttt{load} operation in the kernel code space. The 1\,MB aliasing delay with 7 steps suggests that we can track the address of a kernel memory \texttt{load} by the knowledge of our arbitrary filled store addresses. The blue (No Conflict) line shows the timing when there is no aliasing between the target memory \texttt{load} and the attackers \texttt{store}. Surprisingly, only by filling the store buffer, the system call executes much slower: the normal execution time for \texttt{mincore} should be around $250$ cycles (cyan/No Store). This proof of concept shows that \attack\ can be used to leak information from more privileged contexts, however this is limited only to \texttt{loads} that appear at the beginning of the next context.

\subsection{Negative Result: \attack\ SGX}
\begin{figure}[tp]
  \includegraphics[width=.98\linewidth]{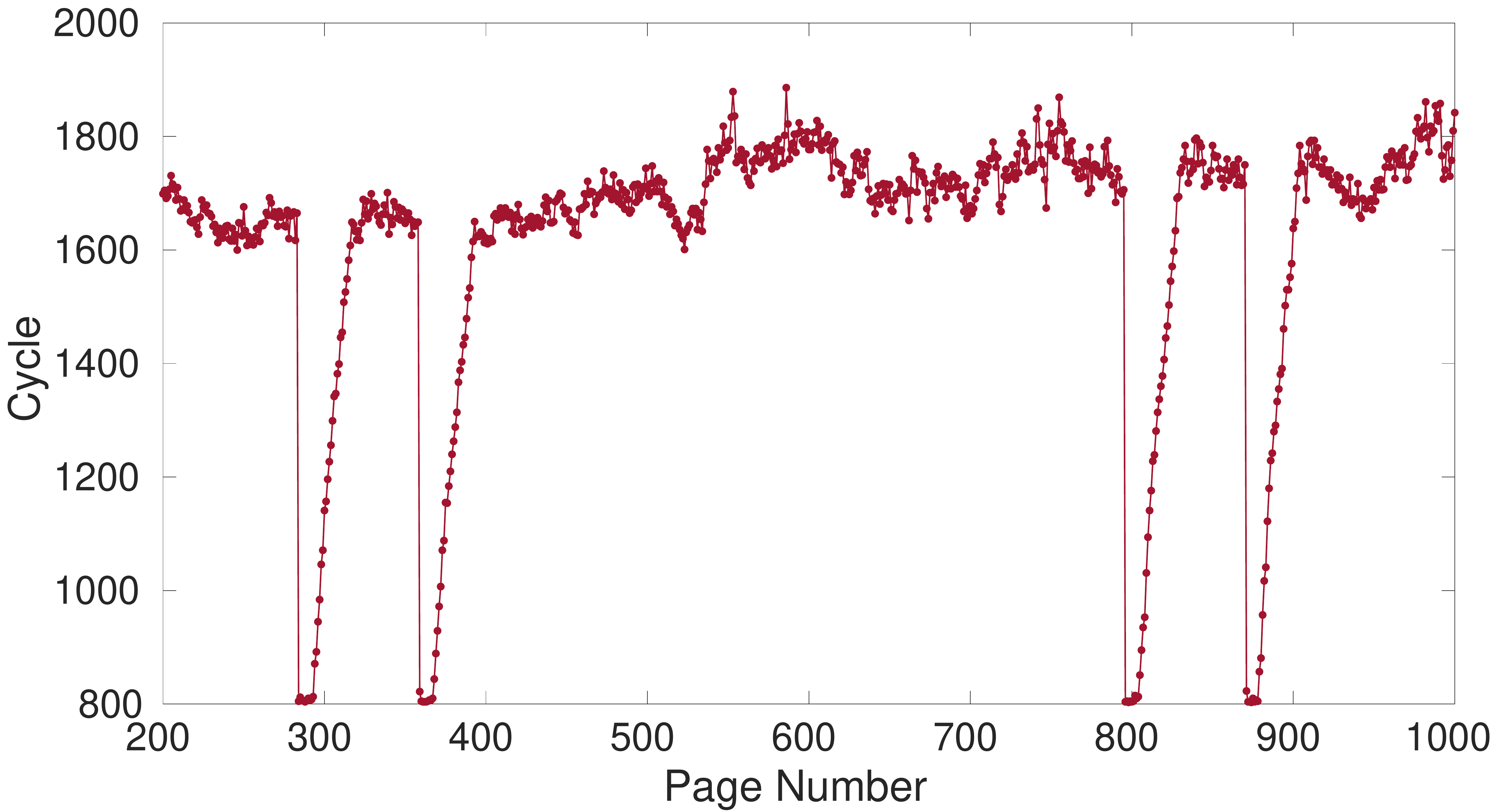}
  \caption{The effect of \attack\ on TLB flush. The execution cycle always increases for 4\,kB aliased addresses, except for some of the virtual pages inside in the store buffer where we observe step-wise hills.}
  \label{fig:flushtlb}
\end{figure}

In this experiment, we try to combine \attack\ with the \textit{CacheZoom}~\cite{moghimi2017cachezoom} approach to create a novel single-threaded side-channel attack against SGX enclaves with high temporal and spatial resolution (4-byte)~\cite{moghimi2018memjam}. We use \textit{SGX-STEP}~\cite{vanbulck2017SGXStep} to precisely interrupt every single instruction. \textit{Nemesis}~\cite{van2018nemesis} shows that the interrupt handler context switch time is dependent on the execution time of the currently running instruction. On our test platform, \textit{Core i7-8650U}, each context switch on an enclave takes about $12000$ cycles to execute. If we fill the store buffer with memory addresses that match the page offset of a \texttt{load} inside the enclave in the interrupt handler, the context switch timing is increased to about $13500$ cycles. While we cannot observe any correlation between the matched 4\,kB or 1\,MB aliased addresses, we do see unexpected periodic downward peaks with a similar step-wise behavior as \attack  (\autoref{fig:flushtlb}).
We later reproduce a similar behavior by running \attack\ before an \texttt{ioctl} routine that flushes the TLB on each call. Intel SGX also performs an implicit TLB flush during each context switch. We can thus infer that the downward peaks occur due to the TLB flush, especially since the addresses for the downward peaks do not have any address correlation with the \texttt{load} address. This suggests that the TLB flush operation itself is affected by \attack. This effect eliminates the opportunity to observe any potential correlation due to the speculative load. As a result, we can not use \attack\ to track memory accesses inside an enclave. Further exploration of the root cause of the TLB flush effect can be carried out as a future work. 

\section{Mitigations}

\medskip\noindent{\bf Software Mitigations} 
The attack exploits the fact that when there is a \texttt{load} instruction after a number of \texttt{store} instructions, the physical address conflict causes a high timing behavior. This happens because of the speculatively executed \texttt{load} before all the \texttt{stores} are finished executing. There is no software mitigation that can completely erase this problem. While the timing behavior can be removed by inserting store fences between the \texttt{loads} and \texttt{stores}, this cannot be enforced to the user's code space, i.e., the user can always leak the physical address information. Another yet less robust approach is to execute other instructions between the \texttt{loads} and \texttt{stores} to decrease the depth of the attack. However, both of the approaches are only applicable to defend against attacks such as the one described in~\autoref{sec:track}. 

As for most attacks on JavaScript, removing accurate timers from the browser would be effective against \attack. Indeed, some timers have been removed or distorted by jitters as a response to attacks~\cite{lipp2018meltdown}. There is however a wide range of timers with varying precision available, and removing all of them seems impractical \cite{schwarz2017fantastic,frigo2018grand}. 

When it is not possible to mitigate the microarchitectural attacks, developers can use dynamic tools to at least detect the presence of such leakage~\cite{chiappetta2016real,zhang2016cloudradar,briongos2018cacheshield}. One of the dynamic approaches is gained by monitoring hardware performance counters in real-time. As explained in~\autoref{sec:hpc}, two of the counters \texttt{Ld\_Blocks\_Partial:Address\_Alias} and \texttt{Cycle\_Activity:Stalls\_Ldm\_Pending} have high correlations with the leakage.

\medskip\noindent{\bf Hardware Mitigations} 
The hardware design for the memory disambiguator may be revised to prevent such physical address leakage, but modifying the speculative behavior may cause performance impacts. For instance, partial address comparison was a design choice for performance. Full address comparison may address this vulnerability, but will also impact performance. Moreover, hardware patches are difficult to be applied to legacy systems and take years to be deployed.

\section{Conclusion}
We introduced \attack, a novel approach for gaining physical address information by exploiting a new information leakage due to speculative execution. To exploit the leakage, we used the speculative load behavior after jamming the store buffer. \attack\ can be executed from user space and requires no special privileges. We exploited the leakage to reveal information on the 8 least significant bits of the physical page number, which are critical for many microarchitectural attacks such as Rowhammer and cache attacks. We analyzed the causes of the discovered leakage in detail and showed how to exploit it to extract physical address information. 

Further, we showed the impact of \attack\ by performing a highly targeted Rowhammer attack in a native user-level environment. We further demonstrated the applicability of \attack\ in sandboxed environments by constructing efficient eviction sets from JavaScript, an extremely restrictive environment that usually does not grant any access to physical addresses. Gaining even partial knowledge of the physical address will make new attack targets feasible in browsers even though JavaScript-enabled attacks are known to be difficult to realize in practice due to the limited nature of the JavaScript environment.  Broadly put, the leakage described in this paper will enable attackers to perform existing attacks more efficiently, or to devise new attacks using the novel knowledge. The source code for \attack\ is available on GitHub\footnote{\url{https://github.com/UzL-ITS/Spoiler}}.

\medskip\noindent{\bf Responsible Disclosure} 
We informed the \textit{Intel Product Security Incident Response Team} (iPSIRT) of our findings. iPSIRT thanked for reporting the issue and for the coordinated disclosure. iPSIRT then released the public advisory and CVE. Here is the time line for the responsible disclosure:

\begin{itemize}[noitemsep]  
\item \textbf{12/01/2018:} We informed our findings to iPSIRT. 
\item \textbf{12/03/2018:} iPSIRT acknowledged the receipt.
\item \textbf{04/09/2019:} iPSIRT released public advisory (INTEL-SA-00238) and assigned CVE (CVE-2019-0162).
\end{itemize}

\section*{Acknowledgments}
We thank Yuval Yarom, our shepherd Eric Wustrow and the anonymous reviewers for their valuable comments for improving the quality of this paper.

This work is supported by U.S. Department of State, Bureau of Educational and Cultural Affairs’ Fulbright Program and National Science Foundation under grant CNS-1618837 and CNS-1814406. We also thank Cloudflare for their generous gift to support our research.

{\footnotesize \bibliographystyle{plain}
\bibliography{main}}

\newpage
\section{Appendix}

\subsection{Tested Hardware Performance Counters}

\begin{table}[ht]
\centering
\begin{tabular}{| c | c |}
\hline
Counters & \textit{Correlation} \\
 \hline  \hline
\scriptsize{UNHALTED\_CORE\_CYCLES}              &  \scriptsize{0.3077}\\
\scriptsize{UNHALTED\_REFERENCE\_CYCLES}         &  \scriptsize{0.1527}\\
\scriptsize{INSTRUCTION\_RETIRED}               &  \scriptsize{0.2718}\\
\scriptsize{INSTRUCTIONS\_RETIRED}              &  \scriptsize{0.2827}\\
\scriptsize{BRANCH\_INSTRUCTIONS\_RETIRED}       &  \scriptsize{0.3143}\\
\scriptsize{MISPREDICTED\_BRANCH\_RETIRED}       &  \scriptsize{0.0872}\\
\scriptsize{CYCLE\_ACTIVITY:CYCLES\_L2\_PENDING}  &  \scriptsize{-0.0234}\\
\scriptsize{CYCLE\_ACTIVITY:STALLS\_LDM\_PENDING} &  \scriptsize{0.9819}\\
\scriptsize{CYCLE\_ACTIVITY:CYCLES\_NO\_EXECUTE}  &  \scriptsize{0.2317}\\
\scriptsize{RESOURCE\_STALLS:ROB}               &  \scriptsize{0}\\
\scriptsize{RESOURCE\_STALLS:SB}                &  \scriptsize{-0.0506}\\
\scriptsize{RESOURCE\_STALLS:RS}                &  \scriptsize{-0.0044}\\
\scriptsize{LD\_BLOCKS\_PARTIAL:ADDRESS\_ALIAS}   &  \scriptsize{-0.9511}\\
\scriptsize{IDQ\_UOPS\_NOT\_DELIVERED}            &  \scriptsize{-0.1455}\\
\scriptsize{IDQ:ALL\_DSB\_CYCLES\_ANY\_UOPS}       &  \scriptsize{0.0332}\\
\scriptsize{ILD\_STALL:IQ\_FULL}                 &  \scriptsize{0.1021}\\
\scriptsize{ITLB\_MISSES:MISS\_CAUSES\_A\_WALK}    &  \scriptsize{0}\\
\scriptsize{TLB\_FLUSH:STLB\_THREAD}             &  \scriptsize{0}\\
\scriptsize{ICACHE:MISSES}                     &  \scriptsize{0}\\
\scriptsize{ICACHE:IFETCH\_STALL}               &  \scriptsize{0}\\
\scriptsize{L1D:REPLACEMENT}                   &  \scriptsize{0.3801}\\
\scriptsize{L2\_DEMAND\_RQSTS:WB\_HIT}            &  \scriptsize{0.2436}\\
\scriptsize{LONGEST\_LAT\_CACHE:MISS}            &  \scriptsize{0.0633}\\
\scriptsize{CYCLE\_ACTIVITY:CYCLES\_L1D\_PENDING} &  \scriptsize{-0.0080}\\
\scriptsize{LOCK\_CYCLES:CACHE\_LOCK\_DURATION}   &  \scriptsize{0}\\
\scriptsize{LOAD\_HIT\_PRE:SW\_PF}                &  \scriptsize{0}\\
\scriptsize{LOAD\_HIT\_PRE:HW\_PF}                &  \scriptsize{0}\\
\scriptsize{MACHINE\_CLEARS:CYCLES}             &  \scriptsize{0}\\
\scriptsize{OFFCORE\_REQUESTS\_BUFFER:SQ\_FULL}   &  \scriptsize{0}\\
\scriptsize{OFFCORE\_REQUESTS:DEMAND\_DATA\_RD}   &  \scriptsize{0.1765}\\ 
 \hline
\end{tabular}
\caption{Counters profiled for correlation test }
\label{table:fullcounter}
\end{table}

\subsection{Row conflict Side Channel} \label{subsec:row-conflict}

\begin{figure}[tp]
  \includegraphics[width=\linewidth]{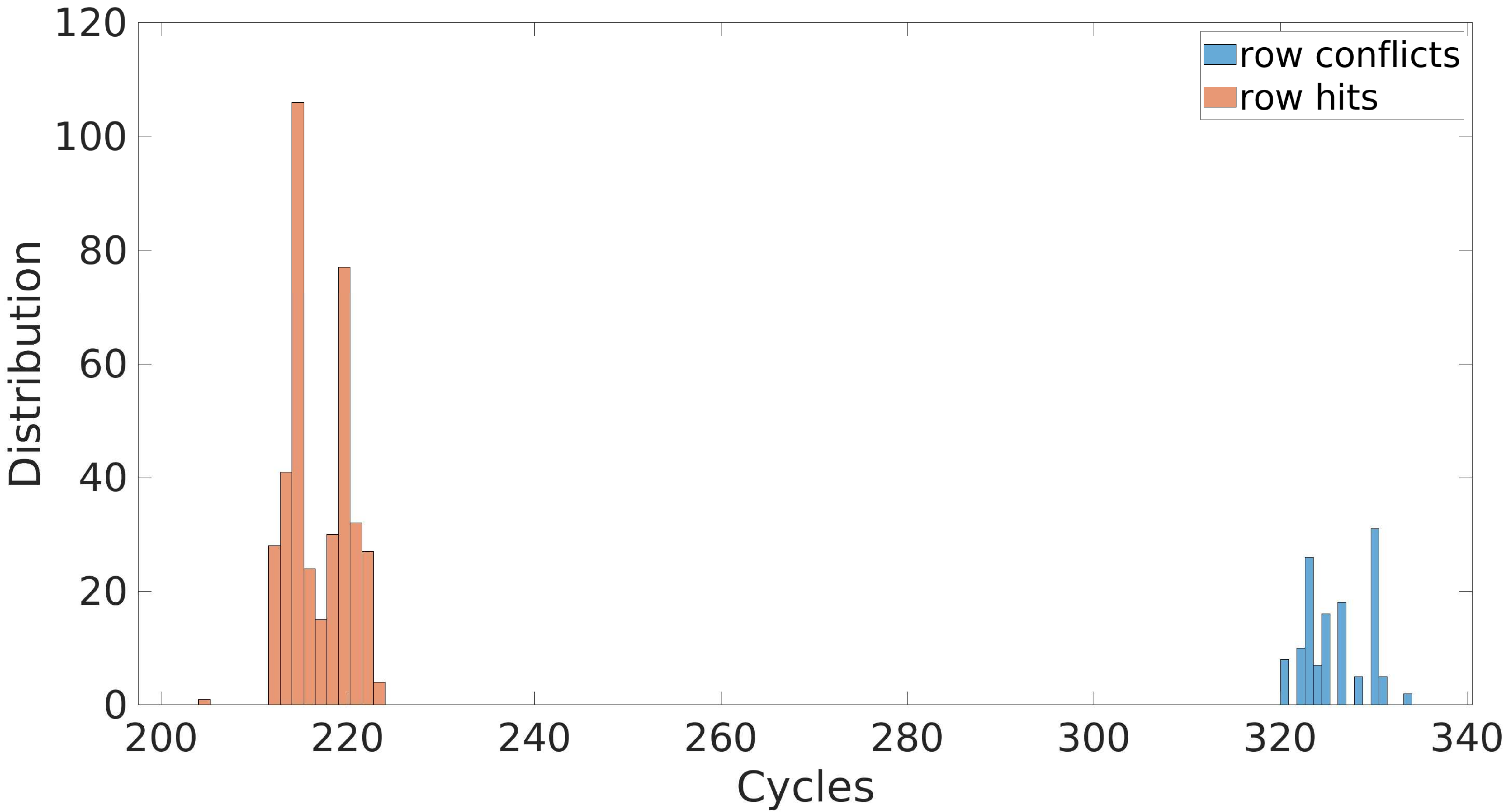}
  \caption{Timings for accessing the aliased virtual addresses (random addresses where 20 LSB of the physical address match). Row hits (orange/low timings) are clearly distinguishable from row conflicts (blue/high timings).}
  \label{fig:rowhits}
\end{figure}

The row conflict side channel retrieves the timing information of the CPU while doing direct accesses (using \texttt{clflush}) from the DRAM. A higher timing indicates that the two addresses are mapped to the same bank in the DRAM because reading an address from the same bank forces the row buffer to copy the previous contents back to the original row and then load the newly accessed data into the row buffer. Whereas, a low timings indicates that two addresses are not in the same bank (not sharing the same row buffer) and are loaded into separate row buffers. \autoref{fig:rowhits} shows a wide gap (around 100 cycles) between row hits and row conflicts. 

\subsection{Memory Utilization and Contiguity} \label{sec:memutil}

The probability of obtaining contiguous memory depends on memory utilization of the system. We conduct an experiment to examine the effect of memory utilization on availability of contiguous memory. In this experiment, 1\,GB memory is allocated. During the experiment, the memory utilization of the system is increased gradually from 20\% to 90\%. We measure the probability of getting the contiguous memory with two methods. The first one is checking the physical frame numbers from \texttt{pagemap} file to look for 520 kB of contiguous memory. The second method is using \attack\ to find the 520 kB of contiguous memory. This 520 kB is required to get three consecutive rows within a bank for a DRAM configuration having 256 kB row offset and 8 kB row size.

\begin{figure}[ht]
  \includegraphics[width=1\linewidth]{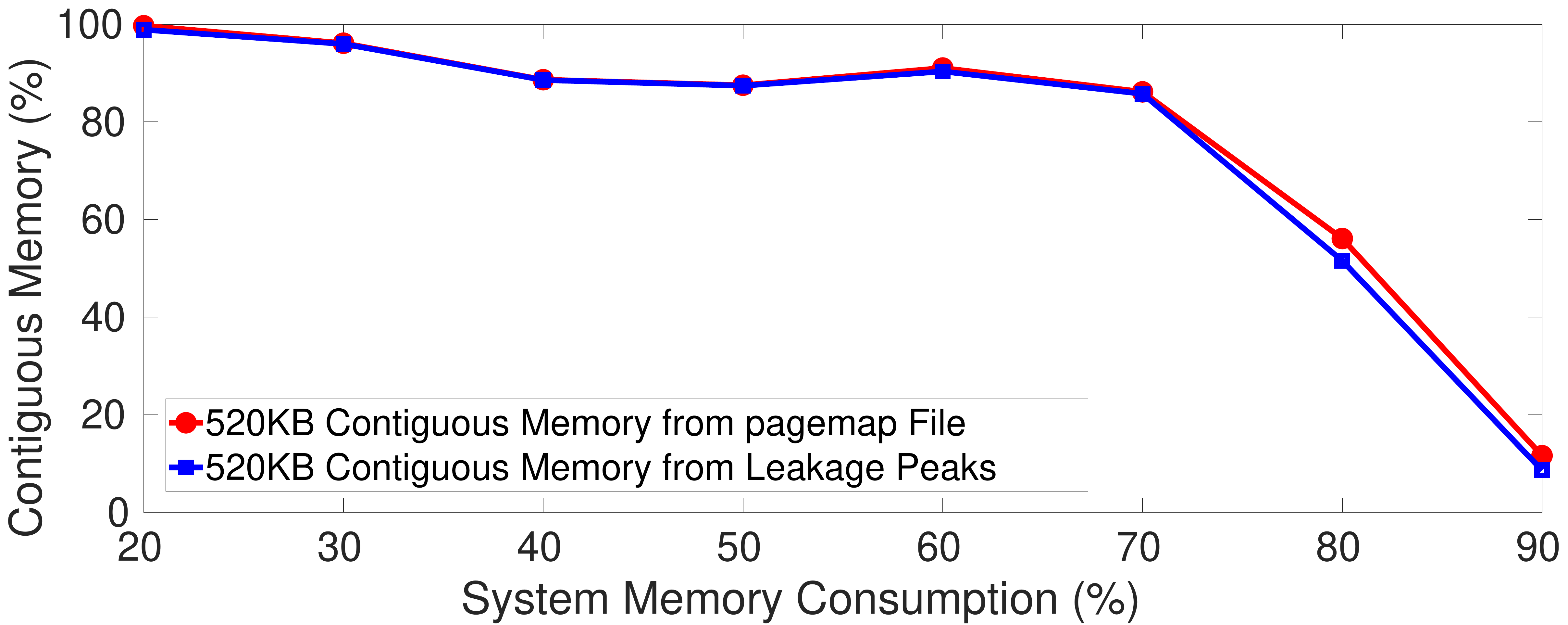}
  \caption{Finding contiguous memory of 520 kB with increasing memory utilization. The overlap between the red and blue plot indicates the high accuracy of the contiguous memory detection capability of \attack\ as verified by the \texttt{pagemap} file.}
  \label{fig:cont_window_inc}
\end{figure}

\begin{figure}[ht]
  \includegraphics[width=1\linewidth]{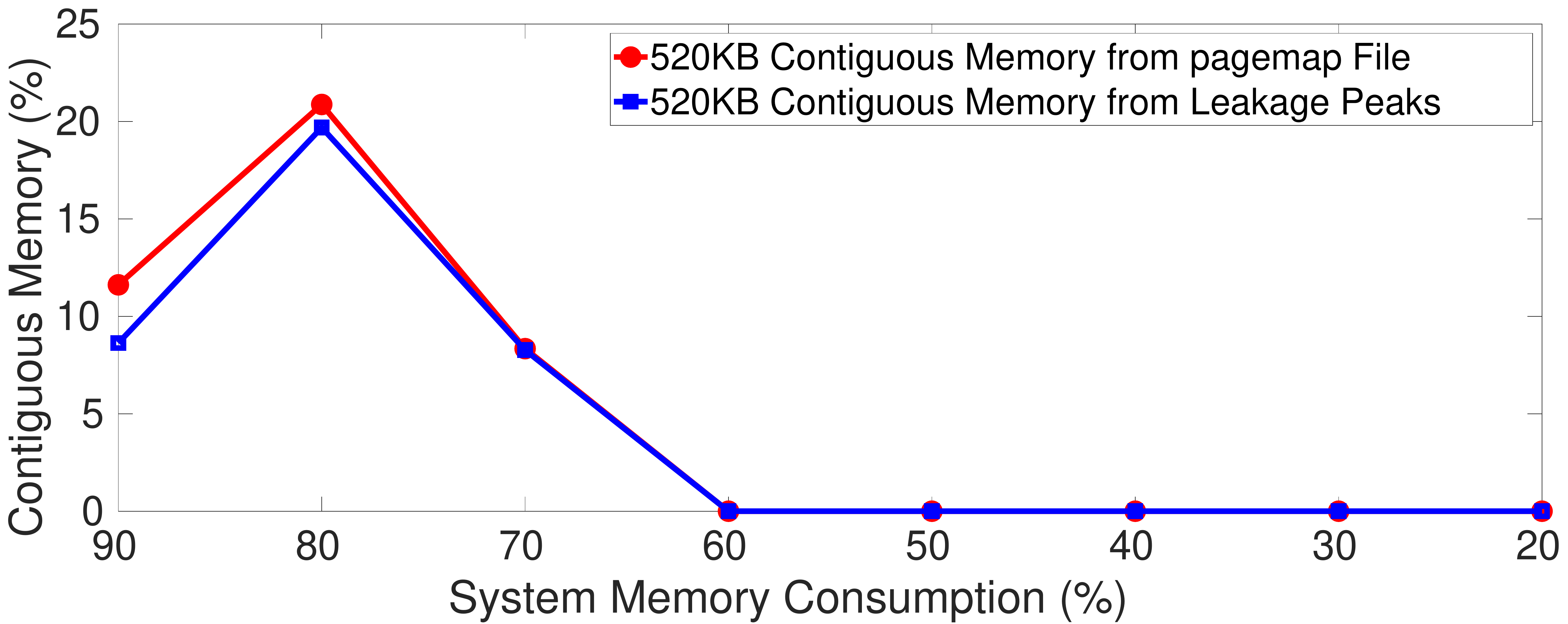}
  \caption{Finding contiguous memory of 520 kB with decreasing memory utilization.}
  \label{fig:cont_window_dec}
\end{figure}

\autoref{fig:cont_window_inc} and~\autoref{fig:cont_window_dec} show that when the memory has been fragmented after intense memory usage, it gets more difficult to allocate a contiguous chunk of memory. Even decreasing the memory usage does not help to get a contiguous block of memory. \autoref{fig:cont_window_dec} depicts that after the memory utilization has been decreased from 70\% to 60\% and so on, there is not enough contiguous memory to mount a successful double-sided Rowhammer attack. Until the machine is restarted, the memory remains fragmented which makes a double-sided Rowhammer attack difficult, especially on targets like high-end servers where restarting is impractical.

The observed behavior can be explained by the \textit{binary buddy allocator} which is responsible for the physical address allocation in the Linux OS~\cite{gorman2004understanding}. This type of allocator is known to fragment memory significantly under certain circumstances~\cite{lloyd1985ontheworst}. The Linux OS uses a \textit{SLAB/SLOB allocator} in order to circumvent the fragmentation problems. However, the allocator only serves the kernel directly. User space memory therefore still suffers from the fragmentation that the buddy allocator introduces. This also means that getting the contiguous memory required for a double-sided Rowhammer attack becomes more difficult if the system under attack has been active for a while. 

%
%
%

\end{document}